\documentclass[pdftex, aps, pra, floatfix, showpacs, twocolumn, superscriptaddress]{revtex4}
\usepackage{times}
\usepackage{graphicx}
\usepackage{amsmath}
\usepackage{amssymb}
\usepackage{latexsym}
\usepackage{epsfig}
\usepackage{bbm}
\usepackage{color}
\definecolor{darkblue}{rgb}{0, 0, 0.7}
\usepackage[colorlinks=true, breaklinks=true, linkcolor=darkblue, citecolor=darkblue, urlcolor=darkblue]{hyperref}

\newcommand{\doi}[2]{\href{http://dx.doi.org/#1}{#2}}
\newcommand{\arxiv}[2]{\href{http://arxiv.org/abs/#1}{#2}}
\newcommand{\urn}[2]{\href{http://nbn-resolving.de/urn/resolver.pl?urn=#1}{#2}}
\newcommand{\web}[2]{\href{#1}{#2}}


\newcommand{\im}{\mathbbm{i}}

\begin{document}

\title{Influence of the particle number on the spin dynamics of ultracold atoms}

\author{Jannes Heinze}
\email{jheinze@physnet.uni-hamburg.de}
\affiliation{I. Institut f\"ur Theoretische Physik, Universit\"at Hamburg, Jungiusstrasse 9, 20355 Hamburg, Germany}

\author{Frank Deuretzbacher}
\email{fdeuretz@itp.uni-hannover.de}
\affiliation{Institut f\"ur Theoretische Physik, Leibniz Universit\"at Hannover, Appelstr. 2, 30167, Hannover, Germany}

\author{Daniela Pfannkuche}
\affiliation{I. Institut f\"ur Theoretische Physik, Universit\"at Hamburg, Jungiusstrasse 9, 20355 Hamburg, Germany}

\begin{abstract}

We study the dependency of the quantum spin dynamics on the particle number in a system of ultracold spin-1 atoms within the single-spatial-mode approximation. We find, for all strengths of the spin-dependent interaction, convergence towards the mean-field dynamics in the thermodynamic limit. The convergence is, however, particularly slow when the spin-changing collisional energy and the quadratic Zeeman energy are equal, i.~e. deviations between quantum and mean-field spin dynamics may be extremely large under these conditions. Our estimates show, that quantum corrections to the mean-field dynamics may play a relevant role in experiments with spinor Bose-Einstein condensates. This is especially the case in the regime of few atoms, which may be accessible in optical lattices. Here, spin dynamics is modulated by a beat note at large magnetic fields due to the significant influence of correlated many-body spin states.

\end{abstract}


\maketitle

\section{Introduction}

The spin degree of freedom of spinor Bose-Einstein condensates (BECs) alows one to explore magnetism of ultracold quantum fluids \cite{Stenger98}, which exhibit intriguing phenomena, like the formation of coreless vortices \cite{Sadler06} and other spin textures \cite{Vengalattore08, Vengalattore09, Kronjaeger09}, and spontaneous symmetry breaking after a quench \cite{Lamacraft07}. Beside the investigation of the ground-state phases \cite{Ho98, Ohmi98, Law98, Koashi00}, the main focus of research is on the study of spin dynamics \cite{Schmaljohann04, Chang04, Kuwamoto04, Widera05, Widera06}. So far most experiments confirmed the mean-field (MF) description, including a nonlinear resonance phenomenon near a critical magnetic field \cite{Chang05, Zhang05, Kronjaeger06, Black07}, which is caused by the interplay of spin-changing collisions and quadratic Zeeman shift. Interestingly, the Hamiltonian of the spin-dependent interatomic interactions appears also in nonlinear quantum optics \cite{Law98}. As a result, spinor BECs allow one to explore, e.~g., four-wave mixing \cite{Kronjaeger06, Goldstein99} and parametric down conversion \cite{Leslie09, Klempt09, Klempt10} with matter waves. Moreover, spin-changing collisions are responsible for the dynamical evolution of squeezed collective spin states and entanglement from uncorrelated (product) states \cite{Sorensen01, Pu00, Duan00, Duan02, Muestecaplioglu02}, which provides a way to overcome the standart quantum limit in precision measurements with matter waves. Those beyond-mean-field correlations can be described by means of the Bogoliubov approximation as long as the quantum fluctuations of the spinor field operator are small \cite{Leslie09, Klempt09, Klempt10, Cui08}.

Correlations limit the validity of the MF approximation. Hence, knowledge of the boundaries of the Gross-Pitaevskii equation (GPE) is of great interest. The validity of the GPE has been proven for weakly interacting spinless bosons in the limit $N \rightarrow \infty$ with $N a$ fixed \cite{Lieb00, Erdos07}, where $N$ is the number of particles and $a$ is the scattering length. We show in this article by means of a numerically exact diagonalization of the effective spin Hamiltonian \cite{Law98, Koashi00, Cui08, Chen09}, that the quantum spin dynamics in the single-mode approximation (SMA) converges towards the MF dynamics in the thermodynamic limit (TDL). This is, interestingly, the case for all strengths of the spin-dependent interaction (within the SMA). The convergence is, however, particularly slow in a regime, where the spin-changing collisional energy and the quadratic Zeeman energy are equal. We determine the validity time of the initial MF dynamics under these conditions, which grows logarithmically with the number of particles. From this we expect that quantum corrections to the MF dynamics may play an important role in experiments with spinor BECs. Additionally, we discuss a beat-note phenomenon in the spin dynamics of few atoms $(N \sim 10)$, which may be observed in deep optical lattices.

The paper is organized as follows. In Sec.~\ref{Sec-methods} we outline the method to calculate the quantum spin dynamics. Afterwards, in Sec.~\ref{Sec-few-atoms}, we discuss few particles. First, in Subsec.~\ref{Subsec-two-atoms} we apply the formalism to two particles to make the method clear and to discuss similarities with the MF dynamics. Then, we discuss the spin dynamics of three particles in Subsec.~\ref{Subsec-three-atoms}, which is modulated by a beat note at large magnetic fields. A similar beat-note phenomenon is recovered for few atoms, which is shown in Subsec.~\ref{Subsec-few-atoms}. In Sec.~\ref{N-particle-quantum-dynamics} we turn to the comparison of the $N$-particle quantum dynamics with the MF dynamics. This is done for two typical initial states. First, in Subsec.~\ref{Subsec-number-state}, we study the initial state, where all atoms are in the $m = 0$ Zeeman sublevel, and then, in Subsec.~\ref{Subsec-transversely-magnetized-state}, we analyze the initial dynamics of the transversely magnetized state. We finally summarize our results in Sec.~\ref{Sec-conclusions}.

\section{Methods}
\label{Sec-methods}

\subsection{Effective spin Hamiltonian}

The two-body interaction between ultracold spin-1 atoms is modeled by a spin-dependent $\delta$ potential \cite{Ho98}
\begin{equation*}
  V_\text{int.} ( \vec r_1 - \vec r_2 ) = \delta ( \vec r_1 - \vec r_2 ) \Bigl( \hbar c_0 + \hbar c_2 \vec f_1 \cdot \vec f_2 \Bigr)
\end{equation*}
with the interaction strengths $c_0$ and $c_2$ and the dimensionless spin-1 matrices $\vec f_i$ of atom $i = 1, 2$. Typically, $c_0$ is one or two orders of magnitude larger than $c_2$. The atoms are confined by a spin-independent trapping potential $V_\text{trap}(\vec r)$. Additionally, a homogeneous magnetic field along the $z$-direction generates the potential
\begin{equation*}
  V_Z = - \hbar p f_z - \hbar q \bigl( 4 - f_z^2 \bigr) ,
\end{equation*}
where $p \propto B$ and $q \propto B^2$ are the coefficients of the linear and quadratic Zeeman energy, respectively. The many-body Hamiltonian is then
\begin{eqnarray*}
  H'' & = & \sum_i \left[ -\frac{\hbar^2}{2 m} \Delta_i + V_\text{trap}(\vec r_i) \right] + \sum_{i < j} \hbar c_0 \delta ( \vec r_i - \vec r_j ) \\
  & & - \sum_i \Bigl[ \hbar p f_{z,i} + \hbar q \bigl( 4 - f_{z,i}^2 \bigr) \Bigr] \\
  & & + \sum_{i < j} \hbar c_2 \delta ( \vec r_i - \vec r_j ) \vec f_i \cdot \vec f_j .
\end{eqnarray*}
The first line contains the spin-independent part of $H''$, which acts only in position space, and the second line contains the Zeeman Hamiltonian, which acts only in spin space. Only the spin-dependent interaction in the third line couples the spin to the motional degrees of freedom.

In the absence of the spin-dependent interaction $(c_2 = 0)$ the states of the ground-state multiplet are of the form
\begin{equation*}
  \psi_0 (\vec r_1, \ldots, \vec r_N) \otimes | \chi_s \rangle
\end{equation*}
with $\psi_0$ being the totally symmetric (nondegenerate) ground state of the spinless problem and with $| \chi_s \rangle$ being an arbitrary totally symmetric $N$-particle spin function \cite{Eisenberg02}. In many experimental situations one can restrict the description to these states \cite{Kronjaeger06, Black07, Widera05, Widera06, Kronjaeger05, Kronjaeger07}, since $|c_2| \ll c_0$. As a consequence, the motion of the atoms is frozen in the ground state $\psi_0$ and the system is essentially zero-dimensional.

Let us assume that we have found the spatial ground state $\psi_0$ and the corresponding energy $E_0$. An integration over the spatial degrees of freedom leads to the effective spin Hamiltonian (the diagonal offset $E_0 - 4 N \hbar q$ is neglected)
\begin{equation*}
  H' = - \hbar p \sum_i f_{z,i} + \hbar q \sum_i f_{z,i}^2 + \hbar g_s \sum_{i < j} \vec f_i \cdot \vec f_j
\end{equation*}
with
\begin{equation*}
  g_s = c_2 \int d \vec r d \vec r_3 \ldots \vec r_N \bigl| \psi_0 (\vec r, \vec r, \vec r_3, \ldots, \vec r_N) \bigr|^2 .
\end{equation*}
The integral is the averaged local pair correlation function $g^{(2)}$ of the ground state $\psi_0$. It can be viewed as the inverse volume of the ground state $g^{(2)} \equiv 1 / V$.

Using the projection operators $N_m = \sum_i | m \rangle_i \langle m |_i ,$ which count the number of particles with polarization $m = +, 0, - ,$ and the relation
\begin{equation*}
  \sum_{i < j} \vec f_i \cdot \vec f_j = \frac{1}{2} \bigl( \vec F^2 - 2 N \bigr) ,
\end{equation*}
where $\vec F^2 = \bigl( \vec f_1 + \ldots + \vec f_N \bigr)^2$ is the square of the total spin, we obtain a more convenient form of the Hamiltonian
\begin{equation*}
  H' = - \hbar p (N_+ - N_-) + \hbar q (N_+ + N_-) + \frac{\hbar g_s}{2} \bigl( \vec F^2 - 2 N \bigr) .
\end{equation*}
The $z$-component of the total spin $F_z = N_+ - N_-$ commutes with the occupation number operators $N_m$ and the Hamiltonian $H'$, which leads to a decomposition of the dynamics into subspaces with different $F_z = M = -N, \ldots, N,$ i.~e.
\begin{equation*}
  \langle N_m \rangle = \sum_M \langle N_m \rangle_M .
\end{equation*}
Therefore, the linear Zeeman energy, which is a constant in each subspace, does not influence the dynamics. In the following we neglect the linear Zeeman energy and set $p = 0$. The remaining Hamiltonian
\begin{equation} \label{Eq-Hamiltonian}
  H = \hbar q (N_+ + N_-) + \frac{\hbar g_s}{2} \bigl( \vec F^2 - 2 N \bigr)
\end{equation}
has a spin-flip symmetry
\begin{equation*}
  H_M = H_{-M} ,
\end{equation*}
since $\vec F^2$ is not changed by a rotation of $180^\circ$ around the $x$-axis and since $N_+ + N_-$ is unaffected if $N_+ \leftrightarrow N_-$.

\subsection{Matrix representation}

The matrix elements of the Hamiltonian (\ref{Eq-Hamiltonian}) are most conveniently calculated within the second quantization formalism. We express the totally symmetric $N$-particle spin functions $| \chi_s \rangle$ by linear combinations of occupation number basis states $| N_+, N_0, N_- \rangle$, which are eigenstates of the occupation number operators $N_m = a_m^\dagger a_m^{}$. The bosonic creation and annihilation operators $a_m^\dagger$, $a_m^{}$ act on these states in the usual way and obey the commutation relations $\bigl[ a_m^{}, a_{m'}^\dagger \bigr] = \delta_{m m'}$ and zero else.

Before we proceed, we change the labeling of the occupation number basis states $| N_+, N_0, N_- \rangle$ in order to simplify the following formulas. We use the set of quantum numbers $(\eta, M, N)$ instead of $(N_+, N_0, N_-)$. Both labels are related to each other through $\eta = N_+ + N_-$, $M = N_+ - N_-$ and $N = N_+ + N_0 + N_-$. We will not explicitly refer to the number of particles $N$, i.~e. $| N_+, N_0, N_- \rangle = | \eta, M \rangle$.

The first summand of (\ref{Eq-Hamiltonian}), the quadratic Zeeman Hamiltonian $H_q = \hbar q (N_+ + N_-)$, is diagonal in the occupation number basis and given by
\begin{equation} \label{Eq-Hamiltonian-a}
  \langle \eta, M | H_q | \eta, M \rangle = \eta \hbar q .
\end{equation}
The second summand, the spin-dependent interaction Hamiltonian $H_s = \frac{\hbar g_s}{2} \bigl( \vec F^2 - 2 N \bigr)$, is tridiagonal. In order to calculate $H_s$ we use the formula
\begin{equation*}
  \vec F^2 = F_z^2 + \frac{1}{2} \Bigl( F_+ F_- + F_- F_+ \Bigr)
\end{equation*}
with the angular momentum creation and annihilation operators $F_{\pm} = F_x \pm \im F_y$, which are given by
\begin{equation*}
  \quad F_{\pm} = \sqrt{2} \, \Bigl( a_{\pm}^\dagger a_0^{} + a_0^\dagger a_{\mp}^{} \Bigr)
\end{equation*}
in dimensionless units. The diagonal elements of $H_s$ are
\begin{equation} \label{Eq-Hamiltonian-b}
  \langle \eta, M | H_s | \eta, M \rangle  =  \frac{\hbar g_s}{2} \bigl[ M^2 - 2 \eta^2 + \eta ( 2N - 1) \bigr]
\end{equation}
and the secondary diagonal elements are
\begin{eqnarray} \label{Eq-Hamiltonian-c}
  \mspace{-50mu} & & \langle \eta + 2, M | H_s | \eta, M \rangle = \langle \eta, M | H_s | \eta + 2, M \rangle = \nonumber \\
  \mspace{-50mu} & & \frac{\hbar g_s}{2} \sqrt{(N - \eta - 1) (N - \eta) (\eta + M + 2) (\eta - M + 2)} \, .
\end{eqnarray}
As discussed before, $H$ decomposes into subblocks, which can be diagonalized independently for each eigenvalue of the total magnetization $M$, since $H$ commutes with $F_z$. The basis states of one subblock are given by
\begin{equation} \label{Eq-subspace-basis}
  \bigl| |M|, M \bigr\rangle, \bigl| |M| + 2, M \bigr\rangle, \ldots, \bigl| \eta_\text{max}, M \bigr\rangle
\end{equation}
with $\eta_\text{max} = N - 1$ or $N$, leading to a total dimension of
\begin{equation} \label{Eq-subspace-dimension}
  \text{dim}(N, M) = \left\lfloor \frac{N - |M|}{2} \right\rfloor + 1
\end{equation}
with the common floor function.

\subsection{Population dynamics}

Once we have determined the eigenstates $| \epsilon \rangle$ and eigenfrequencies $\omega_\epsilon$ of $H$, we can calculate the time evolution of the system, if the initial state $| \psi_i \rangle$ has been specified. The initial state $| \psi_i \rangle$ evolves according to
\begin{equation*}
  | \psi (t) \rangle = \exp \bigl(-\im H t / \hbar \bigr) | \psi_i \rangle ,
\end{equation*}
where the spectral representation of the time evolution operator is given by
\begin{equation} \label{Eq-time-evolution-operator}
  \exp \bigl(-\im H t / \hbar \bigr) = \sum_\epsilon | \epsilon \rangle \langle \epsilon | \exp \bigl(-\im \omega_\epsilon t \bigr) .
\end{equation}
The time evolution of the relative population $n_0 = N_0 / N$ is thus given by
\begin{equation} \label{Eq-time-evolution-of-population}
  n_0(t) = \langle \psi_i |e^{\im H t / \hbar} a_0^\dagger a_0^{} e^{-\im H t / \hbar} | \psi_i \rangle / N .
\end{equation}
The population of the other spin components is completely determined by the conservation of the total number of particles $N$ and of the total magnetization $\langle F_z \rangle$ via
\begin{equation*}
  n_\pm (t) = \frac{1}{2} \Bigl[ 1 \pm \langle F_z \rangle / N - n_0(t) \Bigr] .
\end{equation*}
For the initial states chosen here, $\langle F_z \rangle = 0$. In the following discussion we present only the time evolution of $n_0$ and we neglect any contributions, which are constant in time, since they are determined by the initial state $| \psi_i \rangle$.

\subsection{Initial states}

We will discuss the time evolution of the number state $| \theta_N \rangle = | 0, N, 0 \rangle$, where all the atoms are in the $m = 0$ Zeeman state, and the transversely magnetized state
\begin{equation*}
  | \zeta_N \rangle = \frac{1}{\sqrt{N!}} \biggl( \frac{1}{2} a_+^\dagger + \frac{1}{\sqrt{2}} a_0^\dagger + \frac{1}{2} a_-^\dagger \biggr)^N | 0, 0, 0 \rangle ,
\end{equation*}
where all the spins are pointing into the positive $x$-direction. This state is a superposition of number states from all subspaces
\begin{equation*}
  | \zeta_N \rangle = \sum_{M = -N}^N \sum_{\eta = |M|, \Delta \eta = 2}^{\eta_\text{max}} \chi_\eta^M | \eta, M \rangle
\end{equation*}
with the coefficients
\begin{equation} \label{Eq-coefficients}
  \chi_\eta^M = \biggl( \frac{1}{2} \biggr)^{\frac{\eta + N}{2}} \sqrt{\binom{N}{\eta} \binom{\eta}{\frac{\eta + M}{2}}} \; .
\end{equation}

\subsection{Dimensionless coupling parameter}

The essential parameter, which characterizes the interplay between the quadratic Zeeman energy and the spin-dependent interaction energy, is given by
\begin{equation*}
  K = \frac{2 q}{(2 N - 1) g_s} \rightarrow \frac{q}{c_2 \rho} \quad \text{for large} \; N ,
\end{equation*}
where $\rho = N / V$ is the particle density. $K$ can be positive or negative depending on the sign of $q$ and $c_2$.

\section{Few-atom dynamics}
\label{Sec-few-atoms}

\subsection{Two atoms}
\label{Subsec-two-atoms}

We begin with two atoms to illustrate the method and since typical features of the two-atom dynamics occur also for larger particle numbers and in the MF limit. Two-atom spin dynamics was experimentally investigated in Refs. \cite{Widera05, Widera06} for a system being initially in the number state.

Let us start with the calculation of the matrix elements of the Hamiltonian (\ref{Eq-Hamiltonian}). The matrix $H$ decomposes into 5 submatrices with total magnetization $M = 2, 1, 0, -1, -2$. According to Eq.~(\ref{Eq-subspace-dimension}) the subspace with $M = 0$ has dimension 2 and the others have dimension 1. If one writes down Eq.~(\ref{Eq-time-evolution-of-population}) for an arbitrary initial state, one sees, that only energy differences within the same subspace lead to a sinusoidal oscillation with frequency $\omega_{ij} = | \omega_i - \omega_j |$. Thus, only the $M = 0$ subspace contributes to the time evolution with exactly one frequency $\omega$ and the others lead to constant offset amplitudes, which we will neglect in the following. According to Eq.~(\ref{Eq-subspace-basis}) the two basis states of the $M = 0$ subspace are $| 0, 0 \rangle$ and $| 2, 0 \rangle$. Using Eqs.~(\ref{Eq-Hamiltonian-a}--\ref{Eq-Hamiltonian-c}) we obtain the $2 \times 2$ matrix
\begin{equation} \label{Eq-subspace-matrix-a}
  H^0 = \hbar \begin{pmatrix} 0 & \sqrt{2} g_s \\ \sqrt{2} g_s & 2q - g_s \end{pmatrix} \! .
\end{equation}

Let us consider an arbitrary initial state, which is given by
\begin{equation} \label{Eq-initial-state}
  | \psi_i \rangle = \alpha | 0, 0 \rangle + \beta | 2, 0 \rangle + \text{other terms} .
\end{equation}
$($The other terms are the components of the irrelevant subspaces, like $\gamma | 1, 1 \rangle + \delta | 2, 2 \rangle + \ldots$ and so on. Thus, in general $|\alpha|^2 + |\beta|^2 \leqslant 1$.$)$ From the diagonalization of the matrix (\ref{Eq-subspace-matrix-a}) we obtain the spectral representation of the time evolution operator (\ref{Eq-time-evolution-operator}), which we insert into Eq.~(\ref{Eq-time-evolution-of-population}) together with the initial state (\ref{Eq-initial-state}). The result is an ordinary cosine oscillation
\begin{equation} \label{Eq-two-atom-oscillation}
  n_0' (t) = A \cos (\omega t)
\end{equation}
with the amplitude
\begin{equation} \label{Eq-two-atom-amplitude}
  A = \frac{2}{\omega^2} \Bigl\{ \! 2 \sqrt{2} \Re(\alpha \beta^*) q g_s + g_s^2 \bigl[ 2 ( |\alpha|^2 - |\beta|^2 ) - \sqrt{2} \Re(\alpha \beta^*) \bigr] \! \Bigr\} \! ,
\end{equation}
where $\Re(\gamma)$ denotes the real part of $\gamma$ and $\gamma^*$ is its complex conjugate, and the frequency
\begin{equation} \label{Eq-two-atom-frequency}
  \omega = \sqrt{(2q - g_s)^2 + 8 g_s^2} .
\end{equation}
Time-independent terms have been neglected in Eq.~(\ref{Eq-two-atom-oscillation}), i.~e. $n_0' (t) = n_0 (t) - n_{0, \text{const.}}$.

\begin{figure}[t]
  \includegraphics[width = 0.974\columnwidth]{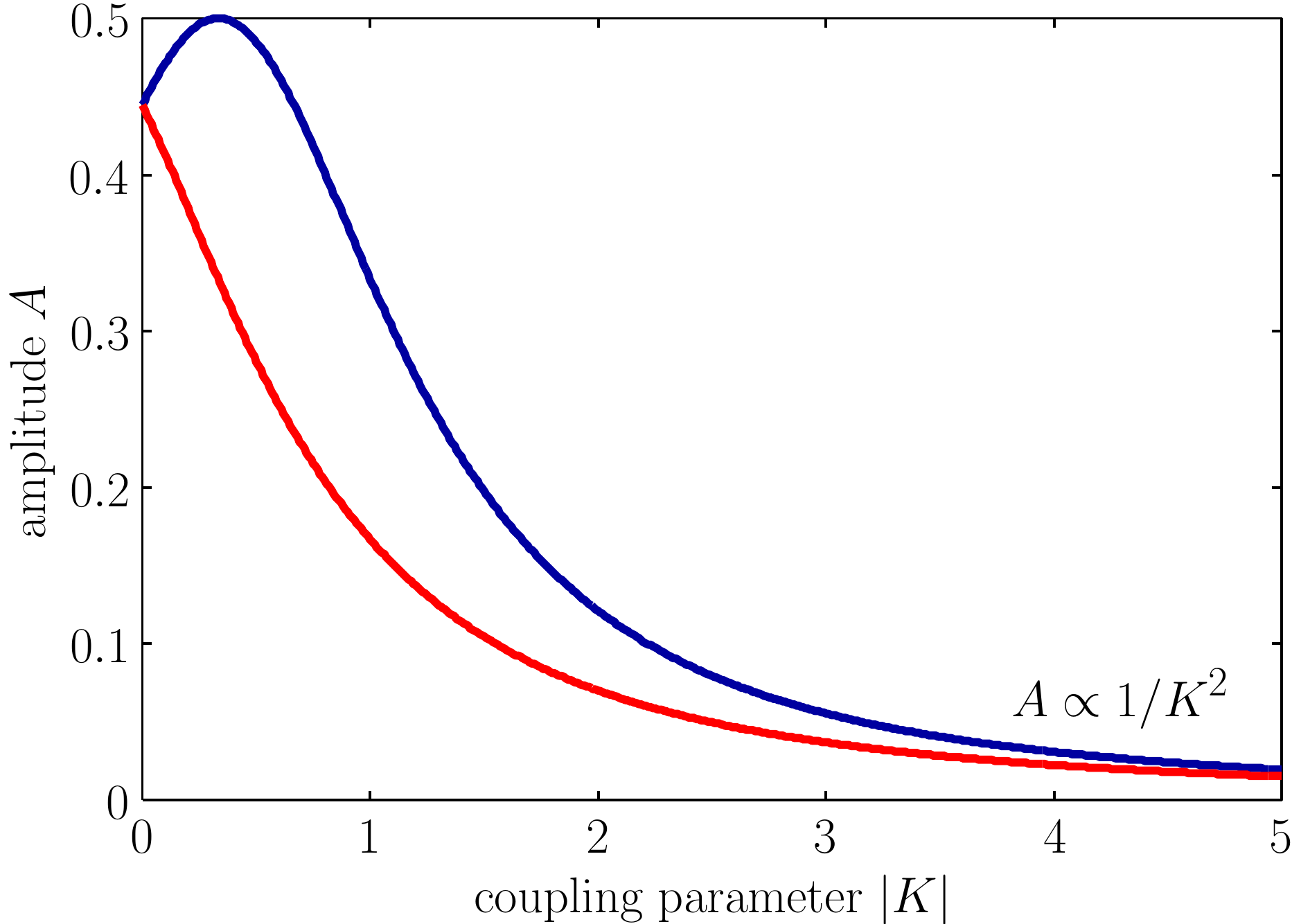}
  \caption{(color online). Two-atom amplitude as a function of $|K|$ for the initially prepared number state, $K < 0$ (red) and $K > 0$ (blue).}
  \label{Fig-two-atom-amplitude-for-theta}
\end{figure}

Let us first discuss the initial state $| \theta_2 \rangle = | 0, 2, 0 \rangle = | 0, 0 \rangle$, i.~e. $(\alpha = 1, \beta = 0)$. In that case, the system undergoes Rabi oscillations \cite{Widera05, Widera06} and the amplitude~(\ref{Eq-two-atom-amplitude}) becomes
\begin{equation} \label{Eq-two-atom-amplitude-for-theta}
  A = \frac{4 g_s^2}{(2 q - g_s)^2 + 8 g_s^2} = \frac{4}{9 - 6 K + 9 K^2} .
\end{equation}
Fig.~\ref{Fig-two-atom-amplitude-for-theta} shows the amplitude $A$ as a function of the coupling parameter $|K|$ for the two cases $K < 0$ (red) and $K > 0$ (blue). For small $|K|$, the Hamiltonian is dominated by the spin-dependent interaction $H_s = \frac{\hbar g_s}{2} \bigl( \vec F^2 - 2 N \bigr)$, and since the number state is not an eigenstate of $H_s$, the system undergoes large oscillations. In the opposite limit of large $|K|$, the Hamiltonian is approximately equal to the quadratic Zeeman energy $H_q = \hbar q (N_+ + N_-)$. Here, the oscillation amplitude converges to zero on the one hand, since the number state is an eigenstate of $H_q$, and on the other hand, since the occupation number operator $n_0 (t) = a_0^\dagger a_0^{} / N$ commutes with $H_q$, and thus becomes a constant of motion.

For the transversely magnetized state $| \zeta_2 \rangle$ we calculate the coefficients $(\alpha = 1 /2, \beta = \sqrt{2} / 4)$ using Eq.~(\ref{Eq-coefficients}). The amplitude~(\ref{Eq-two-atom-amplitude}) takes the form
\begin{equation*}
  A = \frac{q g_s}{(2 q - g_s)^2 + 8 g_s^2} = \frac{K}{6 - 4 K + 6 K^2} .
\end{equation*}
$A$ is negative for $K < 0$, which corresponds to a phase shift of $\pi$ in the cosine function of Eq.~(\ref{Eq-two-atom-oscillation}), and is not in contradiction to the requirement $n_0 (t) > 0$, since $n_{0, \text{const.}} > |A|$.

\begin{figure}[t]
  \includegraphics[width = \columnwidth]{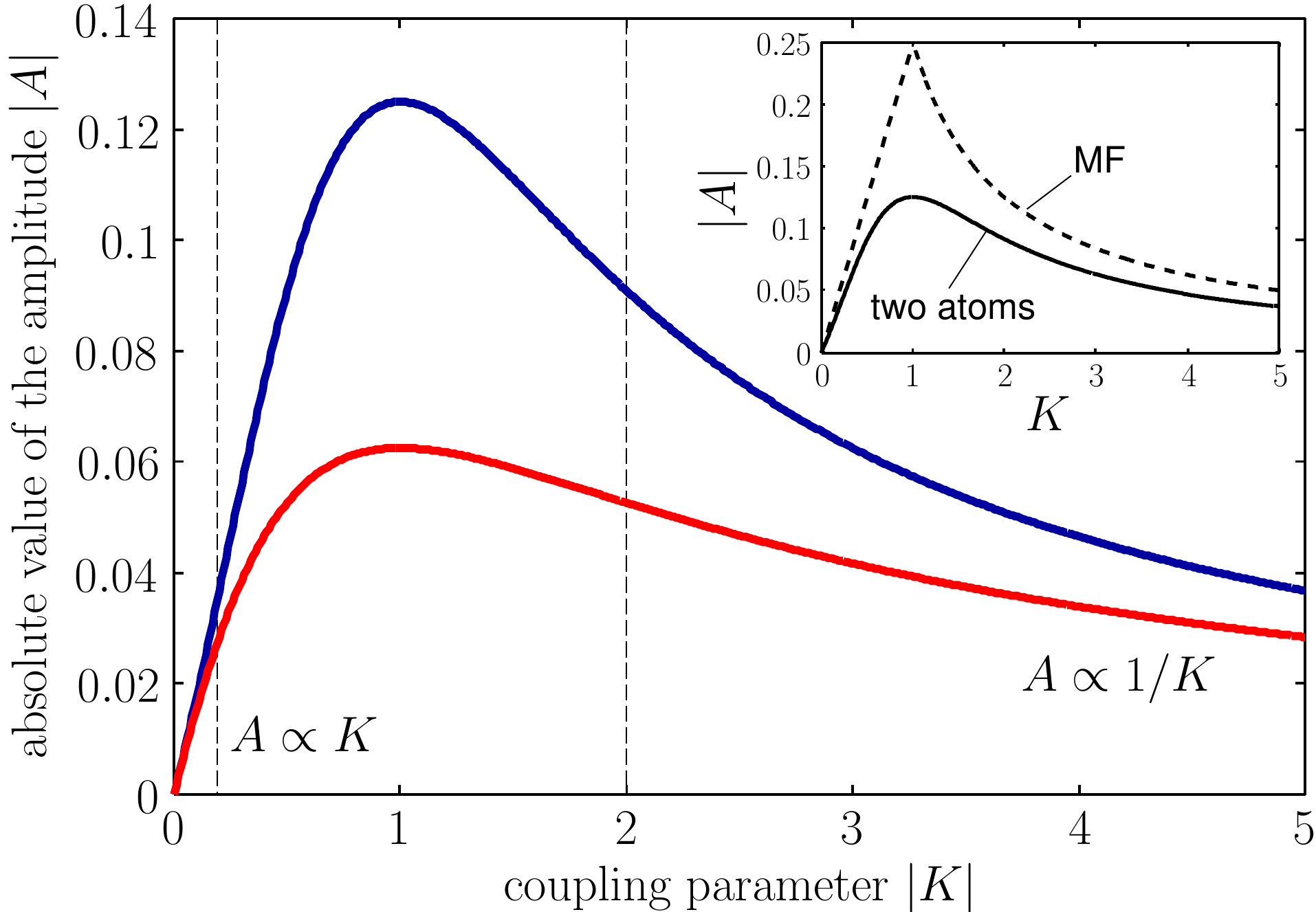}
  \caption{(color online). Absolute value of the two-atom amplitude $|A|$ as a function of $|K|$ for the transversely magnetized state, $K < 0$ (red) and $K > 0$ (blue). Inset: comparison with the MF limit.}
  \label{Fig-two-atom-amplitude-for-zeta}
\end{figure}

Fig.~\ref{Fig-two-atom-amplitude-for-zeta} shows $|A|$ as a function of $|K|$ for $K < 0$ (red) and $K > 0$ (blue). The transversely magnetized state is an eigenstate of the spin-dependent interaction $H_s$ and thus the oscillation amplitude is small for small $|K|$. Again, the amplitude drops down in the opposite limit of large $|K|$, since the occupation number operator $n_0 (t) = a_0^\dagger a_0^{} / N$ commutes with $H_q$ and becomes a constant of motion. Between these limiting cases the amplitude has a maximum, which is located at $|K| = 1$. The MF amplitude (App. \ref{App-mean-field}) has a maximum at the same position and shows the same limiting behavior, which is clear, since the above arguments were independent of the number of particles (inset of Fig.~\ref{Fig-two-atom-amplitude-for-zeta}).

\subsection{Three atoms}
\label{Subsec-three-atoms}

In the case of three atoms, there are 3 subspaces of dimension 2 (those with $M =0, \pm 1$), which contribute to the dynamics. However, since $H$ is symmetric under polarization reflection $M \leftrightarrow -M$, the $M = \pm 1$ subspaces yield the same frequency. The corresponding Hamiltonians are given by
\begin{equation*}
  H^0 = \hbar \begin{pmatrix} 0 & \sqrt{6} g_s \\ \sqrt{6} g_s & 2 q + g_s \end{pmatrix} \! , \;
  H^\pm = \hbar \begin{pmatrix} 0 & 2 g_s \\ 2 g_s & 2 q - 3 g_s \end{pmatrix} \! ,
\end{equation*}
where in $H^\pm$ we have subtracted the offset $\hbar (q + 2 g_s)$ from the diagonal. For the initially prepared number state, the dynamics is restricted to the $M = 0$ subspace, which leads to similar results as in the previously discussed two-atom case.

\begin{figure}[t]
  \includegraphics[width = \columnwidth]{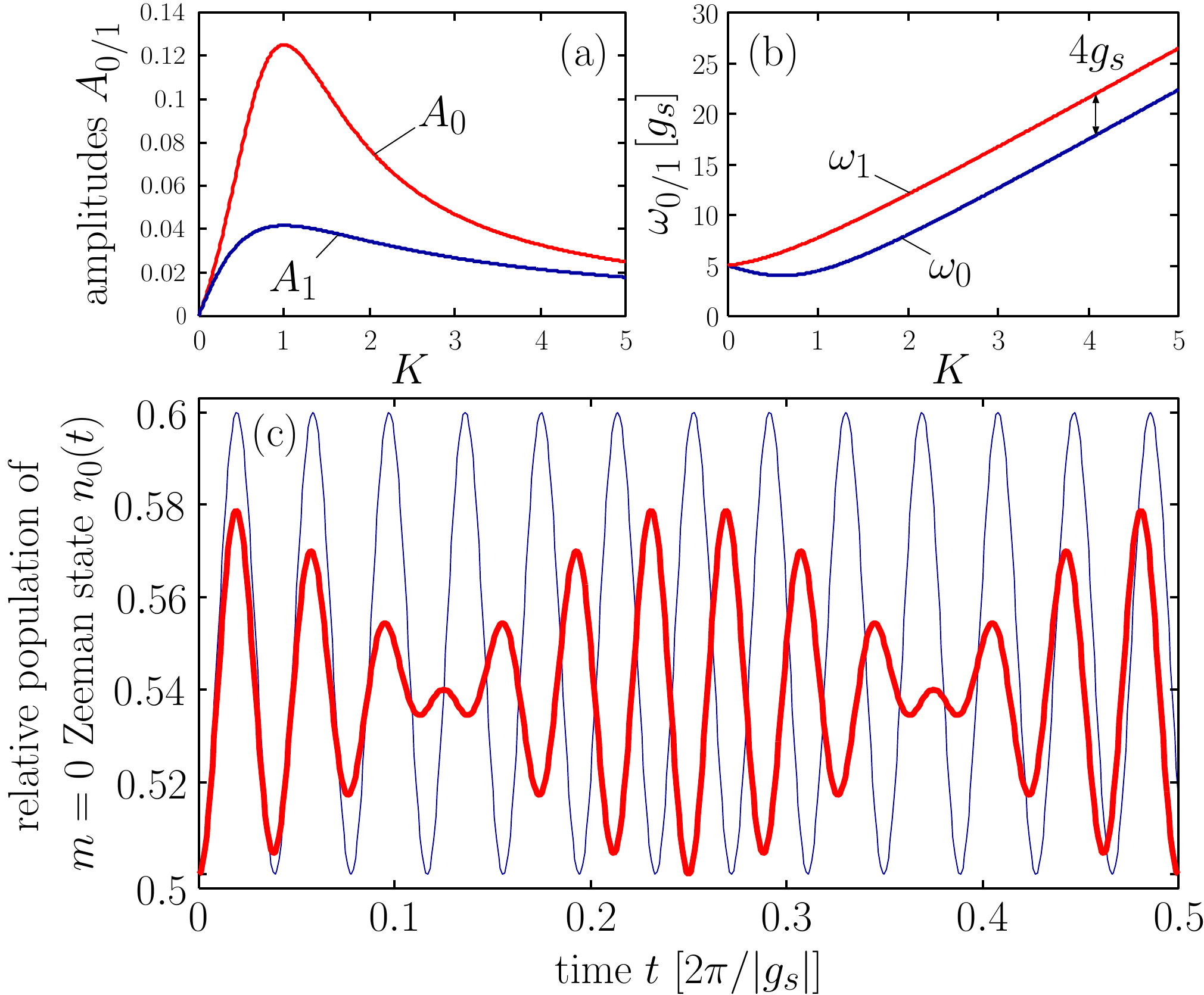}
  \caption{(color online). Evolution of three atoms in the transversely magnetized state. (a) $A_{0/1}$ as a function of $K$. (b) $\omega_{0/1}$ as a function of $K$. (c) Dynamics for $K = -5$ (red) compared to the MF evolution (blue). The oscillation of the $m = 0$ population is modulated by a beat frequency of $2 g_s$, which is absent in the MF limit (App. \ref{App-mean-field}).}
  \label{Fig-three-atom-dynamics}
\end{figure}

By contrast, the transversely magnetized state $| \zeta_3 \rangle$ is distributed over all subspaces. Since all subspaces evolve independently, they can be evaluated in the same way as before and added thereafter. The time evolution of the $m = 0$ population is now given by a sum of two cosines
\begin{equation*}
  n_0'(t) = A_0 \cos (\omega_0 t) + A_1 \cos (\omega_1 t)
\end{equation*}
(constant offsets are neglected) with different amplitudes
\begin{equation*}
  A_{0/1} = q g_s / \omega_{0/1}^2
\end{equation*}
and unequal frequencies
\begin{equation*}
  \omega_0 = \sqrt{(2 q + g_s)^2 + 24 g_s^2} \, , \quad \omega_1 = \sqrt{(2 q - 3 g_s)^2 + 16 g_s^2} \, .
\end{equation*}
The amplitudes $A_0$ and $A_1$ as a function of $K$ show the same behavior as discussed for two atoms [see Fig.~\ref{Fig-three-atom-dynamics}(a)]. The frequencies $\omega_0$ and $\omega_1$ as a function of $K$ are shown in Fig.~\ref{Fig-three-atom-dynamics}(b). At large $K$ they differ by $4 g_s$ so that the oscillation dynamics is modulated by a beat frequency of $2 g_s$:
\begin{eqnarray} \label{Eq-three-atom-beat-note}
  n_0' (t) & \approx & \frac{1}{10 K} \Bigl\{ \cos \bigl[ (2 q + g_s) t \bigr] + \cos \bigl[ (2 q - 3 g_s) t \bigr] \Bigr\} \nonumber \\
  & \approx & \frac{1}{5 K} \cos \bigl[ (2 q - g_s) t \bigr] \cos (2 g_s t) \, .
\end{eqnarray}
Fig.~\ref{Fig-three-atom-dynamics}(c) shows the three-atom dynamics (red) compared to the MF dynamics (blue) at $K = -5$.

\subsection{Dynamics of few atoms for large $K$}
\label{Subsec-few-atoms}

For intermediate values of $K$, the dynamics of few atoms looks rather chaotic, since many frequencies contribute to the evolution. For large $K$, the transversely magnetized initial state shows a fast oscillation, which is modulated by a beat frequency, as in the three-atom case. The limiting dynamics is obtained from a perturbative calculation: For $1 / K \approx 0$, the quadratic Zeeman energy is the dominant part of the Hamiltonian and thus we choose the number states $| N_+, N_0, N_- \rangle$ as a basis. The interaction energy is a small perturbation. In a first step we approximate the eigenstates and energies up to first order in $\lambda = N / (8 K)$. This is separately done for subspaces of $H$ with different magnetization $F_z = M$, since their evolution is decoupled. The result and the representation of the initial state $| \zeta_N \rangle$ (\ref{Eq-coefficients}) is inserted into Eq.~(\ref{Eq-time-evolution-of-population}). After neglecting terms of order $\lambda^2, \lambda^3, \ldots$ one obtains the time evolution
\begin{equation} \label{Eq-many-atom-beat-note}
  n_0' (t) = \frac{N - 1}{2 K (2 N - 1)} \cos \bigl[ (2 q - g_s) t \bigr] \bigl[ \cos (2 g_s t) \bigr]^{N-2} .
\end{equation}

\begin{figure}[t]
  \includegraphics[width = \columnwidth]{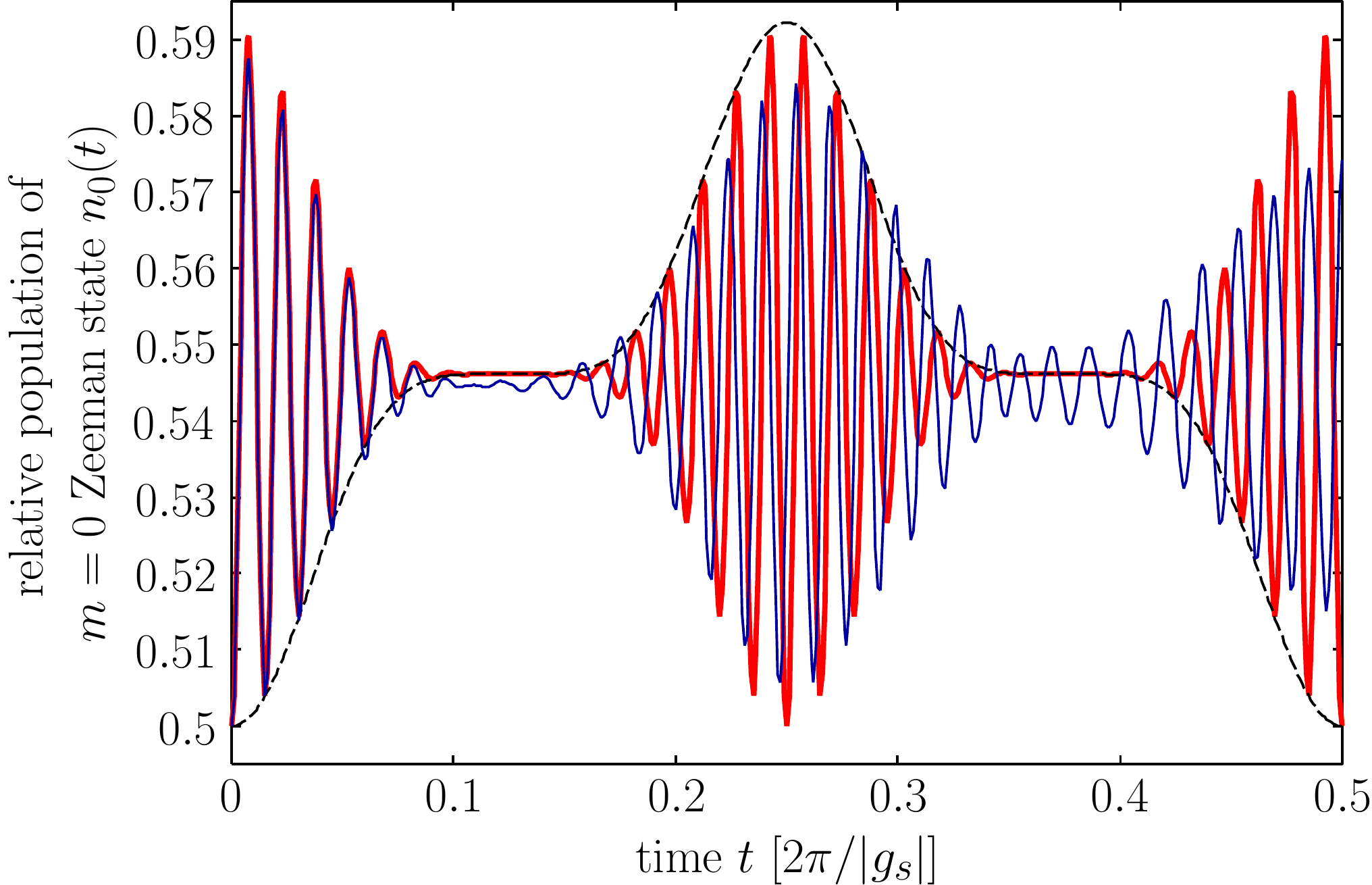}
  \caption{(color online). Evolution of seven atoms in the transversely magnetized state for $K = -5$. The exact numerical calculation (blue) is compared to the first-order perturbative evolution (\ref{Eq-many-atom-beat-note}) (red). The envelope function of (\ref{Eq-many-atom-beat-note}) is drawn as a dashed line.}
  \label{Fig-beat-note}
\end{figure}

The amplitude of the oscillation is proportional to $1 / K$, as in the two-particle and MF limiting cases (Fig.~\ref{Fig-two-atom-amplitude-for-zeta}). The frequency of the fast oscillation, which is approximately given by $\approx 2 q$, is determined by the level spacing of $H_q$, which coincides again with the two-particle (\ref{Eq-two-atom-frequency}) and MF results (\ref{Eq-MF-dynamics}). But different from these two limiting cases, the fast oscillation is modulated by a beat frequency of $2 g_s$.

Fig.~\ref{Fig-beat-note} shows the time evolution of seven atoms at $K = -5$. The beat note is a clear signature that correlated spin states contribute to the dynamics of few atoms. We believe, that the beat frequency can be observed in deep optical lattices with few atoms at each lattice site, similar to the measurements of Refs. \cite{Widera05, Widera06}. The observation is facilitated by the fact, that the beat frequency is independent of the number of particles.

Eq.~(\ref{Eq-many-atom-beat-note}) is valid for rather large particle numbers. We found excellent agreement with the initial evolution of $\lesssim 300$ particles for $K = -5$ from a comparison with the numerical results, although $\lambda \gg 1$ for these parameters.

\section{$N$-particle quantum dynamics vs. mean-field dynamics}
\label{N-particle-quantum-dynamics}

\begin{figure}[t]
  \includegraphics[width = \columnwidth]{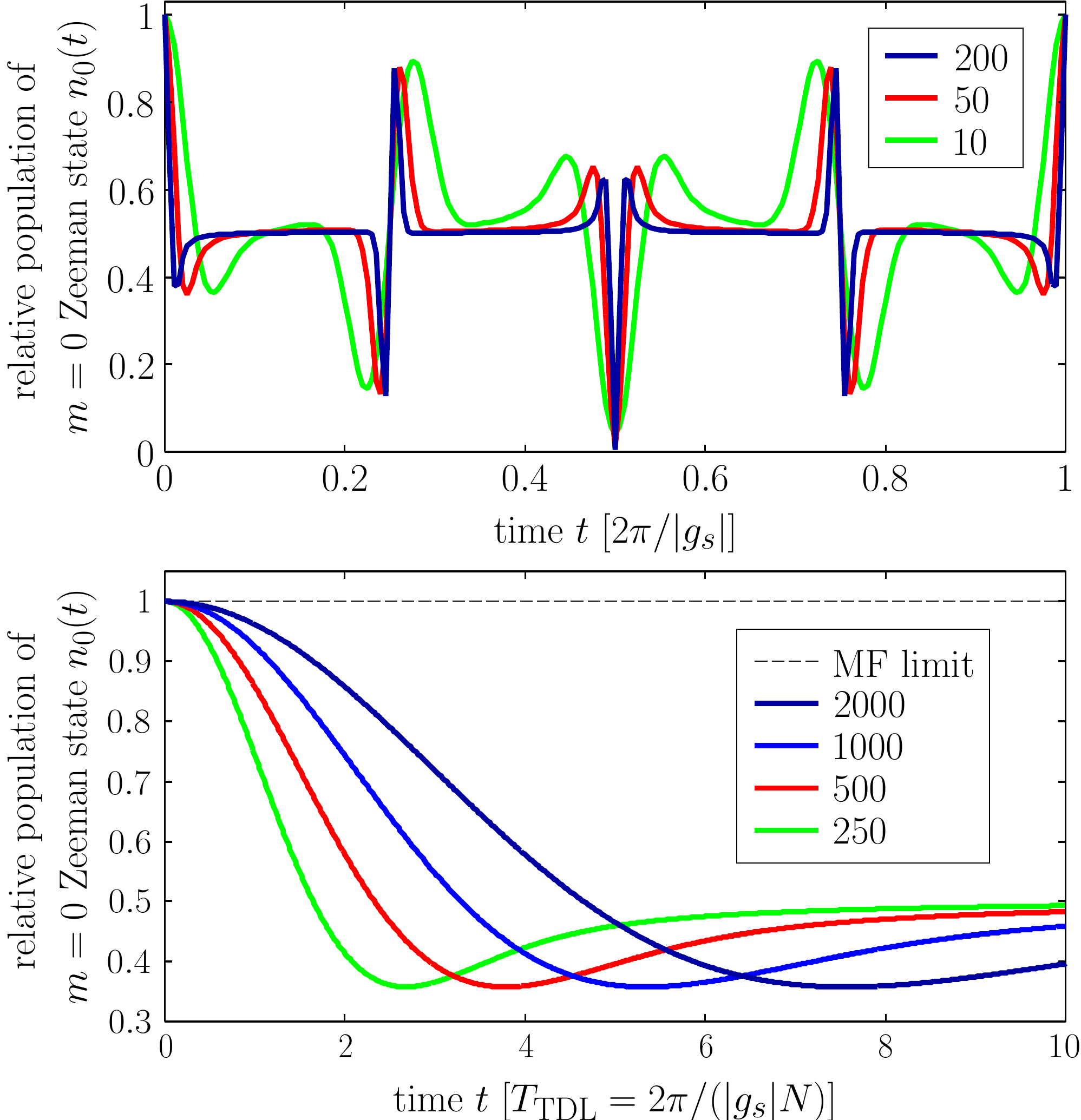}
  \caption{(color online). ({\it top}) Evolution of the number state with $N = 10, 50, 200$ bosons at $K = 0$. The time scale is $2 \pi / |g_s|$. ({\it bottom}) Evolution for $N = 250, 500, 1000, 2000$ bosons at $K = 0$. The time scale is now $T_\text{TDL} = 2 \pi / (|g_s| N) = 2 \pi / (|c_2| \rho)$. In the MF limit, the number state is a steady state, and $n_0 (t) = 1$. With increasing particle number $N$ the population $n_0 (t)$ decreases more slowly on the time scale $T_\text{TDL}$ until it stays constant in the thermodynamic limit.}
  \label{Fig-evolution-of-theta-at-zero-K}
\end{figure}

One can show by means of the large-$N$ method \cite{Yaffe82} that the MF description of the spin system considered here becomes exact in the thermodynamic limit (TDL) \cite{Heinze09}. That means that the time evolution of a coherent initial state is described by the MF equations of motion in the TDL. Further, one of the ground states of the system becomes a coherent state and its energy can be calculated by minimizing the expectation value of $H$ on the MF phase space. Product states (\ref{Eq-product-states}) become identical with coherent states in the limit $N \rightarrow \infty$ and thus the $N$-particle quantum dynamics of the initial states considered here converges towards the MF solution in the TDL. In the following we study the influence of the finite particle number $N$ on the quantum corrections to the MF dynamics.

\subsection{Number state}
\label{Subsec-number-state}

The number state is a steady state of the MF equations of motion for all values of the coupling strength $K$ (App. \ref{App-mean-field}). That makes it particularly useful to study corrections, which go beyond the conventional MF dynamics, since quantum spin fluctuations are strongly amplified in this initial state \cite{Klempt09, Klempt10, Cui08}.

\begin{figure}[t]
  \includegraphics[width = \columnwidth]{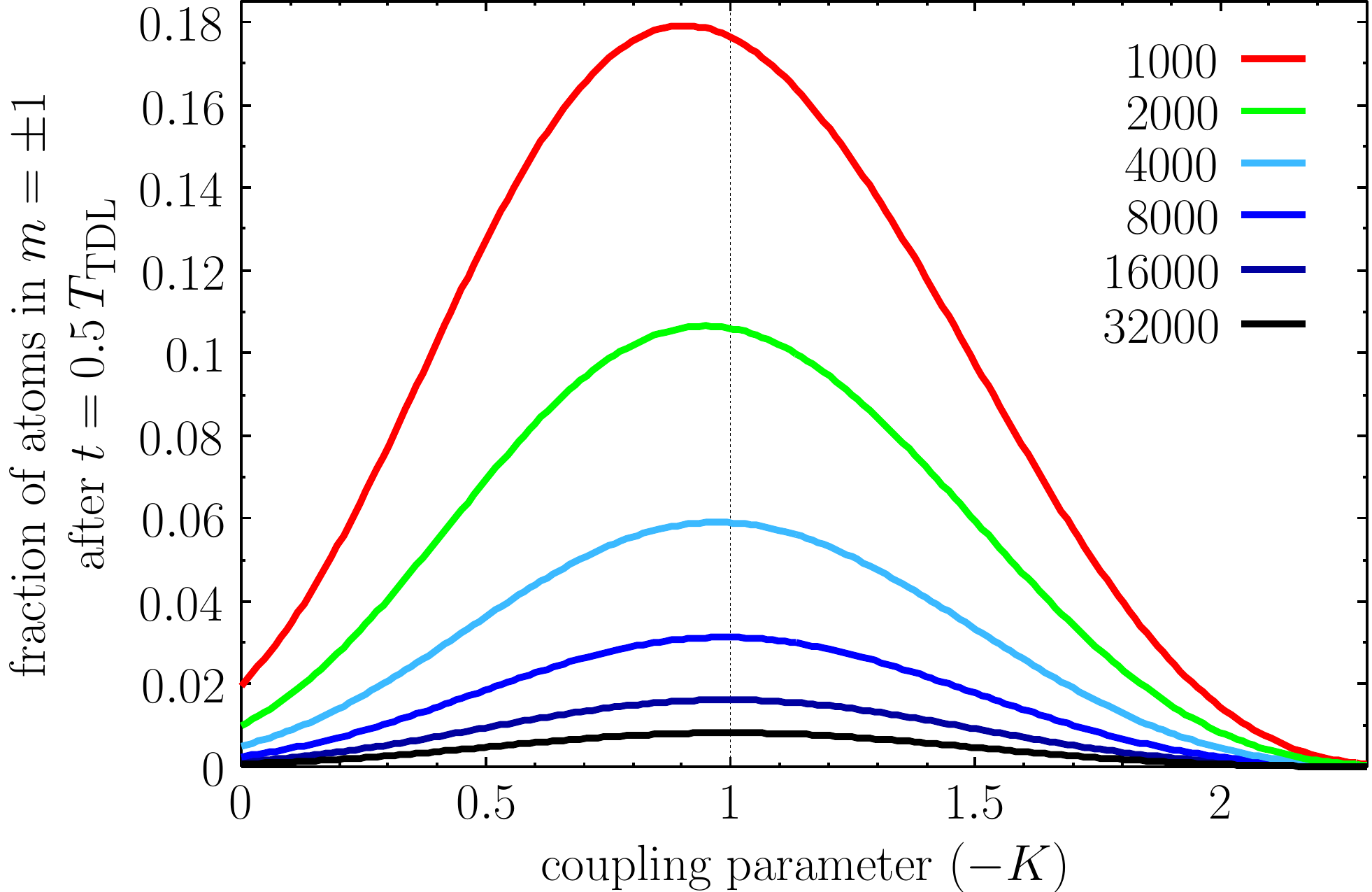}
  \caption{(color online). Relative population of $m = \pm 1$ Zeeman states, $n_\pm = 1 - n_0$, after $t = 0.5 \, T_\text{TDL}$ as a function of $(-K)$. Beyond-MF corrections are largest around $K = -1$ and decrease with increasing particle number.}
  \label{Fig-mf-corrections-as-a-function-of-K}
\end{figure}

\subsubsection{Time evolution for zero $K$}

At $K = 0$, the Hamiltonian consists only of the interaction $H_s$, which has the total-spin states $| F, M \rangle$ as eigenstates. To calculate the $N$-particle quantum dynamics, one needs the representation of the occupation number operator $n_0$ and the initial state $| \theta_N \rangle$ in the total-spin basis ${\{ | F, M \rangle\}}$. The calculation is done in App.~\ref{App-theta-at-zero-K}. The evolution is given by
\begin{equation} \label{Eq-dynamics-of-theta-at-zero-K}
  n_0(t)= \sum_F C_F \cos \bigl[ g_s (2 F + 3) t \bigr] ,
\end{equation}
where the coefficients $C_F$ are approximated by
\begin{equation} \label{Eq-coefficients-of-theta-at-zero-K}
 C_F \approx \frac{F}{N} \exp \left( -\frac{1}{2} \frac{F^2}{N} \right)
\end{equation}
for large particle numbers $N$ (see App.~\ref{App-theta-at-zero-K}). The dynamics is periodic with $2 \pi / |g_s|$, since all the frequencies in (\ref{Eq-dynamics-of-theta-at-zero-K}) are multiples of $g_s$. Fig.~\ref{Fig-evolution-of-theta-at-zero-K}({\it top}) shows a rapid oscillation on the time scale $2 \pi / |g_s|$, which seems to be in contradiction with the MF result, since the number state is a steady state of the MF equations of motion for all $K$. The time scale $2 \pi / |g_s|$, however, becomes infinite in the TDL, since $2 \pi V / |c_2| \rightarrow \infty$ for $V \rightarrow \infty$. Thus, we need a time scale, which stays constant in the TDL. A proper choice is given by
\begin{equation*}
  T_\text{TDL} = 2 \pi / (|g_s| N) = 2 \pi / (|c_2| \rho) ,
\end{equation*}
which is inversely proportional to the spin-dependent interaction energy. By plotting the initial $N$-particle quantum dynamics as a function of $T_\text{TDL}$ [Fig.~\ref{Fig-evolution-of-theta-at-zero-K}({\it bottom})] one sees convergence towards the MF limiting behavior with increasing particle number $N$. The numerical result is confirmed by the analytical formulas: The distribution (\ref{Eq-coefficients-of-theta-at-zero-K}) has a maximum around $F = \sqrt{N}$ and a width proportional to $\sqrt{N}$. Thus, the only cosine oscillation, which contributes to the dynamics (\ref{Eq-dynamics-of-theta-at-zero-K}) has zero frequency in the TDL, since
\begin{equation*}
  2 \sqrt{N} g_s = 2 c_2 \rho / \sqrt{N} \rightarrow 0 \quad (\text{for} \; N \rightarrow \infty) .
\end{equation*}

\subsubsection{Time evolution for large $K$}

A perturbative calculation with the small parameter $\lambda = 1 / (2 K)$ leads to the first-order result
\begin{equation*}
  n_0' (t) = \frac{4 (N - 1)}{K^2 (2 N - 1)^2} \cos \Biggl\{ 2 q t \biggl[ 1 + \frac{2 N - 3}{K (2 N - 1)} \biggr] \Biggr\} .
\end{equation*}
As in the case of two particles the amplitude drops down proportional to $1 / K^2$ [compare with Eq.~(\ref{Eq-two-atom-amplitude-for-theta})] and the frequency of the oscillation converges to $2 q$. Moreover, the amplitude is proportional to $1 / N$ and thus becomes zero in the TDL.

\subsubsection{Beyond-mean-field corrections in dependence of $K$}

We finally analyze the corrections to the MF dynamics in dependence of the coupling parameter $K$. For that reason, we have numerically calculated the fraction of atoms in the $m = \pm 1$ Zeeman states, $n_\pm (t) = 1 - n_0 (t)$, after a fixed time $t = 0.5 \, T_\text{TDL}$ as a function of $(-K)$. Fig.~\ref{Fig-mf-corrections-as-a-function-of-K} shows a resonant enhancement of the $m = \pm 1$ population around $K = -1$. For $N = 1000$ the relative population of the $m = \pm 1$ Zeeman states is largest and the resonance maximum is slightly below $|K| = 1$. With increasing particle number $N$ the resonance maximum decreases and its position converges to $K = -1$.

\begin{figure}[t]
  \includegraphics[width = \columnwidth]{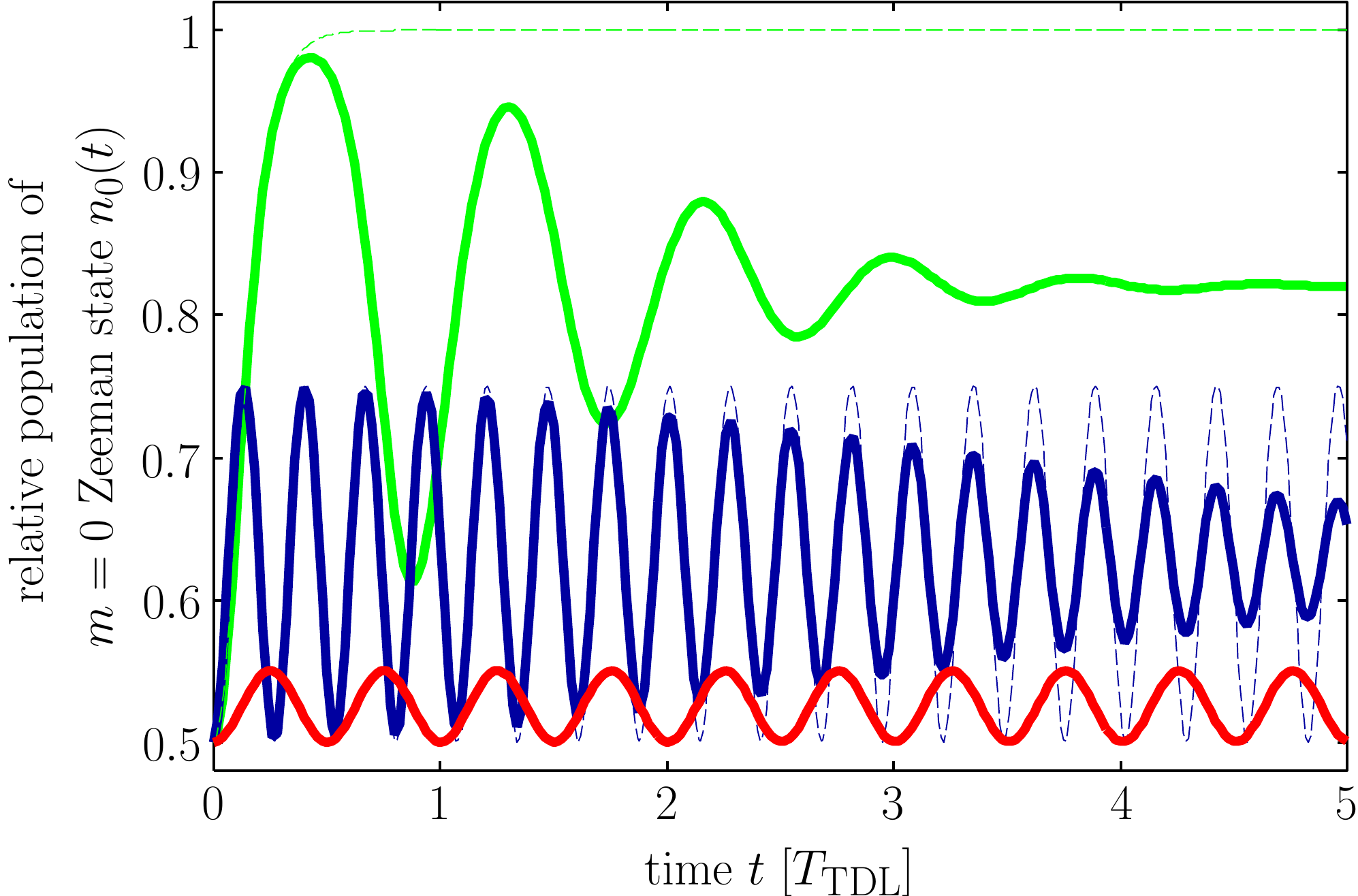}
  \caption{(color online). Evolution of the $m = 0$ population for 2000 particles in the transversely magnetized state for different coupling strengths: $K = -0.1$ (red), $K = -1$ (green) and $K = -2$ (blue). The corresponding MF evolution is drawn as a dashed line.}
  \label{Fig-evolution-of-zeta-for-various-K}
\end{figure}

Beyond-MF corrections were analyzed in a similar way in a recent experiment \cite{Klempt09, Klempt10} and explained within the Bogoliubov theory. Different from here, two resonances have been observed there, which is due to the restricted motion of the atoms to the ground state in our approach. However, on the first resonance, the corresponding Bogoliubov mode has a similar shape than the MF ground state so that our method is applicable there. Indeed, in a large trap, both approaches lead to the same value for the position of the first resonance, namely $|q / (c_2 \rho)| = 1$.

\subsection{Transversely magnetized state}
\label{Subsec-transversely-magnetized-state}

\subsubsection{Time evolution for three different values of $K$}

The transversely magnetized state shows a rich dynamics in the MF limit \cite{Kronjaeger05, Kronjaeger06}. Fig.~\ref{Fig-evolution-of-zeta-for-various-K} shows the initial evolution of $N = 2000$ atoms for three different coupling strengths. We compare the exact $N$-particle quantum dynamics (solid line) to the corresponding MF evolution (dashed line).

At small $|K|$ (red) one sees no difference between the two solutions within the time ${5 \, T_\text{TDL}}$. The evolution is an ordinary cosine oscillation with amplitude $A = K / 4$ and frequency $\omega = 2 c_2 \rho$ ($\rightarrow$ period $T_\text{TDL} / 2$), which is determined by the spin-dependent interaction energy (App. \ref{App-mean-field}).

In the opposite limit of large $|K|$, the MF evolution (blue dashed) is given by a cosine oscillation with amplitude $A = 1 / (4 K)$ and frequency $\omega = 2 q$ $[ \rightarrow$ period $T_\text{TDL} / (2 K) ]$. The exact $N$-particle quantum dynamics (blue solid) oscillates with the same fast frequency of $2 q$. Moreover, the fast oscillation is modulated by the envelope function of Eq. (\ref{Eq-many-atom-beat-note}). The beat note vanishes in the TDL, since
\begin{equation*}
  \bigl[ \cos (2 g_s t) \bigr]^{N-2} \approx 1 - \frac{2 (c_2 \rho t)^2}{N} .
\end{equation*}
Despite the beat note, both oscillations coincide quite well for times, which are smaller than $2 \, T_\text{TDL}$.

\begin{figure}[t]
  \includegraphics[width=\columnwidth]{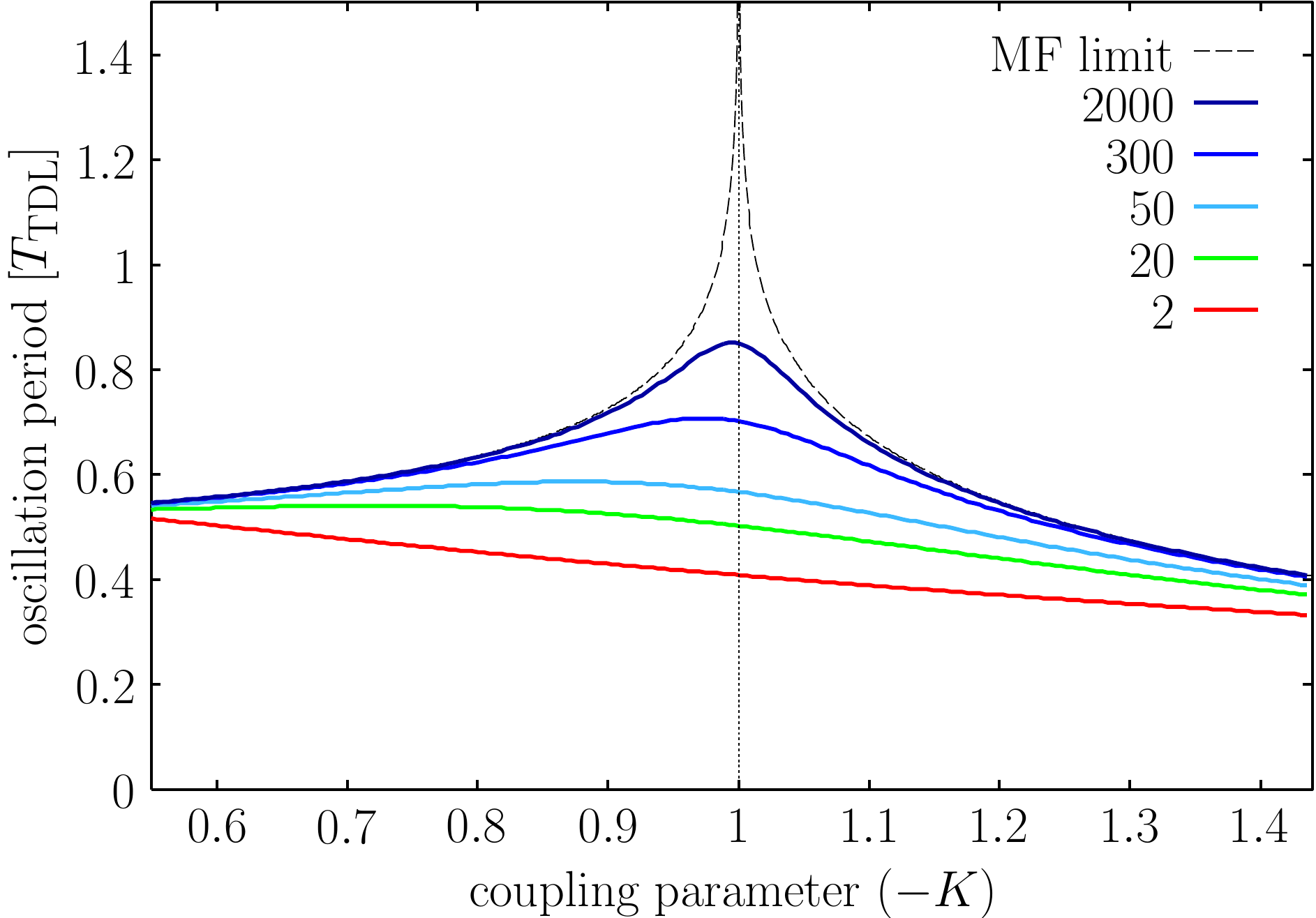}
  \caption{(color online). Oscillation period of the initial evolution in dependence of $(-K)$. For small and large $|K|$, the $N$-particle and MF period are in good agreement. In the vicinity of $K = -1$, the period is strongly enhanced in systems with large particle numbers. The enhancement of the oscillation period is absent in small systems with less than 20 particles.}
  \label{Fig-oscillation-period}
\end{figure}

At ${K = -1}$, the relative population of the $m = 0$ Zeeman state converges towards 1 in the MF limit (green dashed). The amplitude becomes maximal with $A = 1 / 4$ and the oscillation period diverges. The period of the exact $N$-particle quantum evolution (green solid) is enhanced, but still finite. Moreover, the green solid curve does not converge towards 1, but rather oscillates around 0.82. Note, that the evolution is coherent and exhibits a revival although the initial dynamics seems to be strongly damped and to converge towards a constant value. At ${K = -1}$, one observes the largest deviations from the MF dynamics. The initial evolutions coincide only for times, which are smaller than $T_\text{TDL} / 2$. A similar observation was made in the experiment \cite{Kronjaeger06, Black07}.

\subsubsection{Beyond-mean-field corrections in dependence of $K$}

Fig.~\ref{Fig-oscillation-period} shows plots of the oscillation period against the coupling strength $(-K)$ for different particle numbers and the MF limit. In the MF limit, the oscillation period was obtained from the analytical solution (App. \ref{App-mean-field}), while in the other cases, the period is 2 times the position of the first maximum of the initial oscillation.

One sees that the $N$-particle oscillation period coincides quite well with the MF period in the limit of small and large $|K|$, for all particle numbers. In the resonance region around $K = -1$, the deviations are quite large for small particle numbers; see the curves for $N = 2, 20$ and 50. With increasing particle number, the $N$-particle period coincides almost everywhere with the MF limit, except in a small region around $K = -1$, which shrinks to zero in the TDL; see the curves for $N = 300$ and 2000. However, exactly on resonance at $K = -1$, the deviation from the MF limit is always infinitely large for finite particle numbers $N$.

\begin{figure}[t]
  \includegraphics[width=\columnwidth]{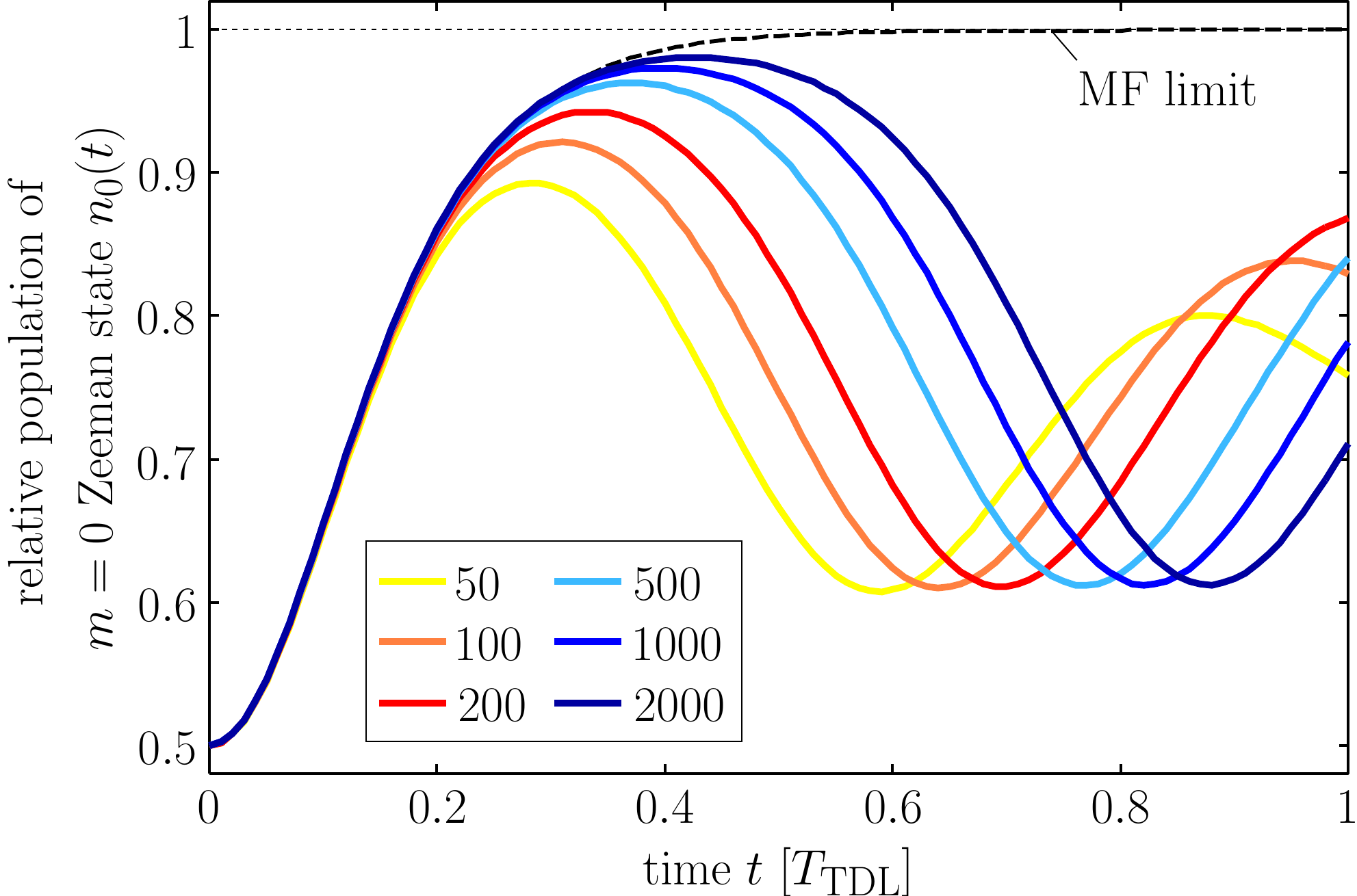}
  \caption{(color online). Initial evolution of the relative population of the $m = 0$ Zeeman state on resonance $(K = -1)$ for different particle numbers $N$ compared to the MF limit.}
  \label{Fig-first-oscillation-on-resonance}
\end{figure}

The enhancement of the oscillation period is a typical feature of the spin systems with large particle numbers. One clearly sees the first occurence of a weak maximum for $N = 20$, which becomes rather pronounced for $N = 2000$. For particle numbers smaller than $\approx 20$, the maximum is absent. One further sees that the position of the resonance maximum rapidly approaches 1 from below with increasing particle number $N$.

The enhancement of the oscillation period has been experimentally observed in an antiferromagnetic spin-2 $^{87}$Rb BEC \cite{Kronjaeger06} and, for a similar initial state, in a spin-1 $^{23}$Na BEC \cite{Black07}. The large differences between the exact $N$-particle quantum dynamics and the MF evolution close to the critical coupling strength at $|K| = 1$ due to finite-$N$ corrections may be relevant for studies of quantum chaos \cite{Kronjaeger08, Weiss08}.

\subsubsection{Limit for the validity of the mean-field dynamics}

We conclude from the previous discussion, that the convergence to the MF limiting dynamics in systems with a finite number of particles is slowest at $|K| = 1$; see Figs.~\ref{Fig-mf-corrections-as-a-function-of-K}--\ref{Fig-oscillation-period}. Hence, we study now, how fast the oscillation period converges towards infinity at the critical value of $|K| = 1$, since this worst-case scenario provides us with the smallest upper bound for the validity of the MF approximation.

\begin{figure}[t]
  \includegraphics[width=\columnwidth]{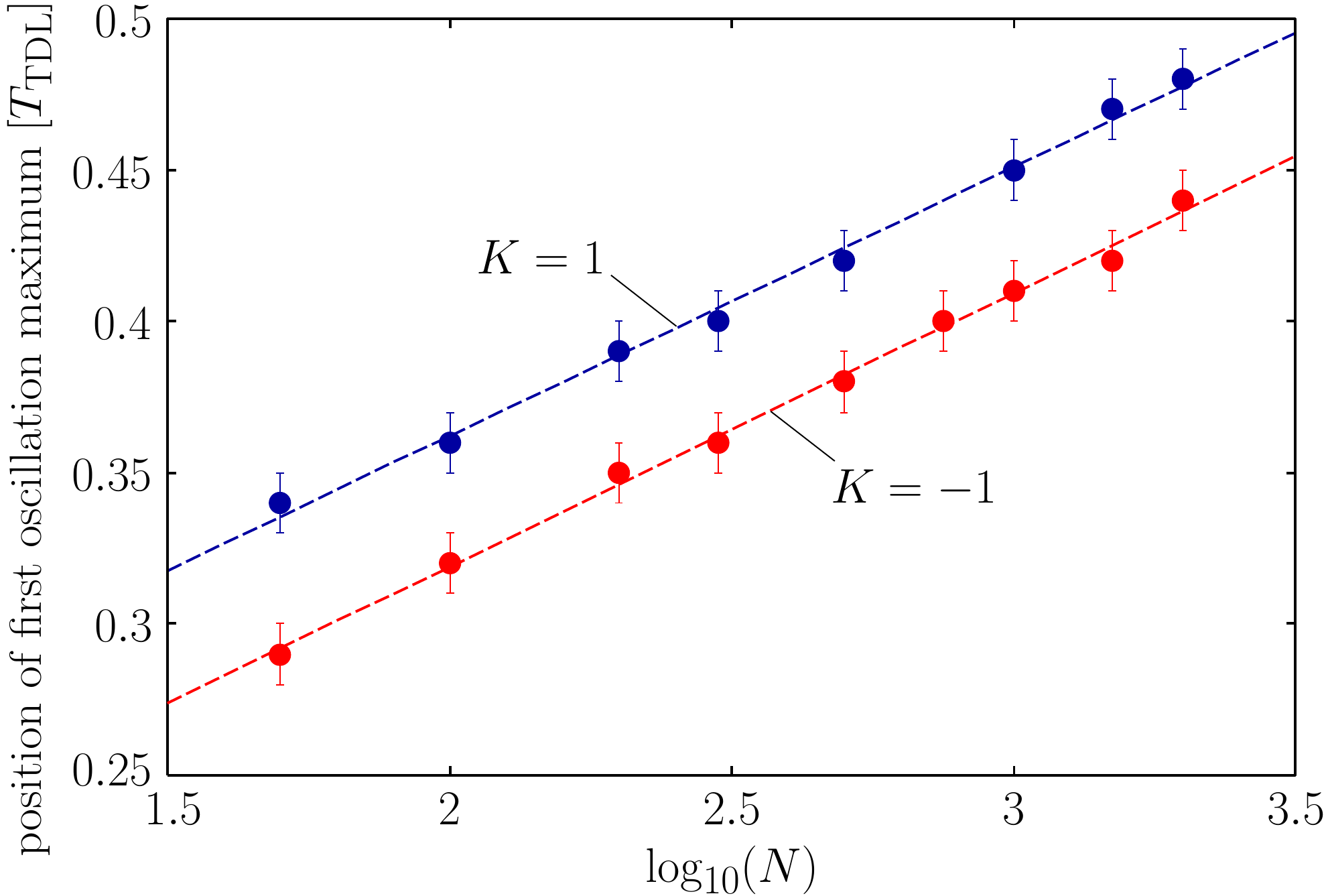}
  \caption{(color online). Position of the first oscillation maximum against the number of particles $N$ on resonance $(|K| = 1)$.}
  \label{Fig-period-of-first-oscillation-on-resonance}
\end{figure}

Fig.~\ref{Fig-first-oscillation-on-resonance} shows the initial evolution of the relative population of the $m = 0$ Zeeman state at $K = -1$ for $N = 50 - 2000$ particles. As expected, the initial evolution coincides with the MF dynamics for longer times, when the particle number is increased. One sees that the $N$-particle quantum dynamics is close to the MF evolution for times, which are smaller than the position of the first oscillation maximum.

Hence, we plot in Fig.~\ref{Fig-period-of-first-oscillation-on-resonance} the position of the first oscillation maximum against the number of particles on a double logarithmic scale. One finds a logarithmic dependency of the validity time $T_K(N)$ of the MF approximation on the number of particles
\begin{eqnarray*}
  T_{K = -1} (N) & = & T_\text{TDL} \left[ 0.13 + 0.09 \log_{10} (N) \right] \\
  T_{K = 1} (N) & = & T_\text{TDL} \left[ 0.18 + 0.09 \log_{10} (N) \right] .
\end{eqnarray*}
Thus, for a spin-1 BEC of $10^6$ atoms and a negative coupling parameter, $K < 0$, one finds the MF dynamics to be valid for times, which are smaller than $\approx 0.67 \, T_\text{TDL}$.

\section{Conclusions and outlook}
\label{Sec-conclusions}

We studied the quantum spin dynamics of ultracold spin-1 atoms within the single-mode approximation. This was done on the basis of an exact diagonalization of the effective many-body spin Hamiltonian \cite{Law98}. The chosen method allowed us to discuss the crossover from few atoms to small condensates within one framework.

Our numerical calculations showed convergence of the quantum spin dynamics towards the mean-field dynamics in the thermodynamic limit for all strengths of the spin-dependent interaction. Moreover we showed that quantum corrections to the mean-field dynamics are particularly large at the critical value of the coupling parameter, where the spin-changing collisional energy equals the quadratic Zeeman energy. For the state, where all the atoms are in the $m = 0$ sublevel, we compared the initial quantum spin dynamics to results obtained from a Bogoliubov approximation \cite{Klempt09, Klempt10, Cui08}. For the transversely magnetized state we estimated the validity time of the initial mean-field dynamics at the critical value of the coupling parameter, which grows logarithmically with the number of particles. From this estimate we conclude that quantum corrections to the mean-field dynamics may play an important role in experiments with spinor BECs. It would be interesting to study systematically the dependency of the spin dynamics on the particle number in future experiments.

Quantum corrections to the mean-field spin dynamics are particularly large in the regime of few atoms. Here, many-body spin correlations lead to a beat-note phenomenon at large magnetic fields. The regime of few atoms $(N \sim 10)$ should be accessible in deep optical lattices.

Our results may be relevant for the study of quantum chaos. It was shown \cite{Kronjaeger08} that the strong nonlinear behavior of the mean-field equations at the critical value of the coupling parameter leads to classical chaos in the spin dynamics of spin-2 atoms. But exactly at this point, the finite-$N$ quantum dynamics largely deviates from the classical limit. It would be interesting to study this aspect systematically, similar to the work in Ref. \cite{Weiss08} on a periodically driven double-well system.

\begin{acknowledgments}

The authors wish to thank K. Sengstock, K. Bongs and F. Lechermann for fruitful discussions. F. D. acknowledges funding by the DFG (SFB407, QUEST) and the ESF (EUROQUASAR), and is grateful for the hospitality of S. M. Reimann at the Division of Mathematical Physics, LTH, Lund University, Sweden. This work has been performed within the Excellence Cluster Frontiers in Quantum Photon science, which is supported by the Joachim Hertz Stiftung.

\end{acknowledgments}

\begin{appendix}

\section{Mean-field dynamics}
\label{App-mean-field}

One assumes in the MF approach, that the system is in a product state of the form
\begin{equation} \label{Eq-product-states}
  \bigl( \alpha_+ | + \rangle + \alpha_0 | 0 \rangle + \alpha_- | - \rangle \bigr)^{\otimes N}
\end{equation}
with $\alpha_m$ being complex numbers, which are normalized according to $\sum_m |\alpha_m|^2 = 1$. For the states (\ref{Eq-product-states}) and the Hamiltonian (\ref{Eq-Hamiltonian}), one derives the MF equations of motion \cite{Kronjaeger05}
\begin{eqnarray} \label{Eq-MF-equations-of-motion}
  \im \partial_t \alpha_+^{} & = & g_s N \Big( A^* \alpha_0^{} + \langle f_z \rangle \, \alpha_+^{} \Big) + q \alpha_+^{} \nonumber \\
  \im \partial_t \alpha_0^{} & = & g_s N \Big( A \, \alpha_+^{} + A^* \alpha_-^{} \Big) \\
  \im \partial_t \alpha_-^{} & = & g_s N \Big( A \, \alpha_0^{} - \langle f_z \rangle \, \alpha_-^{} \Big) + q \alpha_-^{} \, , \nonumber
\end{eqnarray}
where we have defined $A = \langle f_+ \rangle / \sqrt{2} = (\alpha_+^* \alpha_0 + \alpha_0^* \alpha_-)$ and $\langle f_z \rangle = (|\alpha_+|^2 - |\alpha_-|^2)$.

The number state $|0, N, 0 \rangle = | 0 \rangle^{\otimes N}$ is a steady state of (\ref{Eq-MF-equations-of-motion}), since
\begin{equation*}
  \alpha_0 (t) = 1 , \quad \alpha_{\pm} (t) = 0
\end{equation*}
is a solution of the MF equations of motion (\ref{Eq-MF-equations-of-motion}) for all values of $g_s$ and $q$. Thus, this state shows no population dynamics in the MF limit and $n_0 (t) = |\alpha_0 (t)|^2 = 1$.

The time evolution of the transversely magnetized state
\begin{equation*}
  | \zeta_N^y \rangle = \bigl( - 1 / 2 \, | + \rangle - \im / \sqrt{2} \, | 0 \rangle + 1 / 2 \, | - \rangle \bigr)^{\otimes N}
\end{equation*}
is calculated in Refs.~\cite{Kronjaeger05, Kronjaeger07}. In this state, all the spins are pointing into the positive $y$-direction. The solution of the MF Eqs. (\ref{Eq-MF-equations-of-motion}) is given in terms of Jacobi elliptic functions \cite{Bronstein}
\begin{eqnarray*}
  \alpha_{\pm} (t) & = & \mp \frac{s}{2} \left[ \frac{\text{cn}_{k} (\frac{q t}{2}) \text{dn}_{k} (\frac{q t}{2})}{1 - k \text{sn}^2_{k} (\frac{q t}{2})} - \frac{\im (1 + k) \text{sn}_{k} (\frac{q t}{2})}{1 + k \text{sn}^2_{k} (\frac{q t}{2})} \right] , \\
  \alpha_0 (t) & = & \frac{s}{\sqrt{2}} \left[ \frac{(1 - k) \text{sn}_{k} (\frac{q t}{2})}{1 - k \text{sn}^2_{k} (\frac{q t}{2})} - \frac{\im \text{cn}_{k} (\frac{q t}{2}) \text{dn}_{k} (\frac{q t}{2})}{1 + k \text{sn}^2_{k} (\frac{q t}{2})} \right] ,
\end{eqnarray*}
where $s = \exp (-\im (g_s N - q) t / 2)$ and $k = 1 / K$. For the spin populations $n_m (t) = |\alpha_m (t)|^2$ the solution simplifies to
\begin{eqnarray} \label{Eq-MF-dynamics}
  n_0 (t) & = & \bigl[ 1 - k \, \text{sn}_k^2 (q t) \bigr] / 2 , \\
  n_{\pm} (t) & = & \bigl[ 1 + k \, \text{sn}_k^2 (q t) \bigr] / 4 \, . \nonumber
\end{eqnarray}
These solutions are also valid for the initial state $| \zeta_N \rangle$, which is used here, since the evolution of the relative populations $n_m$ is unaffected by rotations around the $z$-axis.

For small $k = 1 / K$, one can approximate $\text{sn}_k (x) \approx \sin (x)$ and Eq. (\ref{Eq-MF-dynamics}) becomes
\begin{equation*}
  n_0 (t) \approx \biggl( \frac{1}{2} - \frac{1}{4 K} \biggr) + \frac{1}{4 K} \cos (2 q t) \qquad (\text{large} \, K) .
\end{equation*}
That means, the evolution is a cosine oscillation with amplitude $A = 1 / (4 K)$ and frequency $\omega = 2 q$.

For large $k = 1 / K$, we approximate $\text{sn}_k (x) \approx \sin (k x) / k$, which leads to
\begin{equation*}
  n_0 (t) \approx \biggl( \frac{1}{2} - \frac{K}{4} \biggr) + \frac{K}{4} \cos (2 c_2 \rho t) \qquad (\text{small} \, K) .
\end{equation*}
That means, in the interaction dominated regime, the oscillation amplitude is $A = K / 4$ and the frequency is $\omega = 2 c_2 \rho$.

At $|K| = 1$, the evolution becomes aperiodic and the relative $m = 0$ population asymptotically approaches 1, i.~e. $n_0 (t) \rightarrow 1$ for $t \rightarrow \infty$. Here, the dynamics exhibits a maximum of the amplitude and the oscillation period diverges.

\section{Quantum dynamics of the number state at zero $K$}
\label{App-theta-at-zero-K}

In the following, we derive the population dynamics of the number state at zero $K$. In this limiting regime, the Hamiltonian consists only of the interaction $H_s$, which has the eigenbasis ${\{| F, M \rangle\}}$. The occupation number operator $N_0$ and the initial state $| \theta_N \rangle$ need to be expressed in this basis to calculate the dynamics. The expansion of $| \theta_N \rangle$ is given by
\begin{equation} \label{App-theta-Eq-1}
  | \theta_N \rangle = |0, N, 0 \rangle = \sum_{F = F_\text{min}, \Delta F = 2}^N \chi_F |F, 0 \rangle .
\end{equation}
Since $| \theta_N \rangle$ has the $F_z$ eigenvalue $M = 0$, it is a superposition of the states $|F, M = 0 \rangle$. Due to symmetry reasons, the summation runs over $F = F_\text{min}, F_\text{min} + 2, \ldots, N$ with $F_\text{min} = 0$ or 1 if $N$ is even or odd, respectively \cite{Eisenberg02}. The coefficients $\chi_F$ are determined later. By inserting the expansion (\ref{App-theta-Eq-1}) into Eq.~(\ref{Eq-time-evolution-of-population}) the evolution becomes
\begin{equation} \label{App-theta-Eq-2}
  n_0 (t) = \sum_{F, F'} \chi_F \chi_{F'} \langle F, 0 | N_0 | F', 0 \rangle \cos \bigl[ (\omega_F - \omega_{F'}) t \bigr] / N
\end{equation}
with the frequencies $\omega_F = g_s \bigl[ F (F + 1) - 2 N \bigr]$. The matrix elements $\langle F, 0 | N_0 | F', 0 \rangle$ are calculated in the following.

The occupation number operator $N_0$ can be written in terms of spherical tensor operators
\begin{equation*}
  N_0 = \frac{1}{3} N - \sqrt{\frac{2}{3}} T_0^{(2)} ,
\end{equation*}
where $T_0^{(2)}$ is the zeroth component of the one-particle spherical tensor operator $T^{(2)}_q$ of rank 2. Its five components are
\begin{eqnarray}
  T_{\pm 2}^{(2)} & = & a_\pm^\dagger a_\mp^{} \label{App-theta-Eq-3a} \\
  T_{\pm 1}^{(2)} & = & \frac{1}{\sqrt{2}} \Bigl( a_0^\dagger a_\mp - a_\pm^\dagger a_0 \Bigr) \nonumber \\
  T_0^{(2)} & = & \frac{1}{\sqrt{6}} \Bigl( a_+^\dagger a_+ - 2 a_0^\dagger a_0 + a_-^\dagger a_- \Bigr) . \label{App-theta-Eq-3b}
\end{eqnarray}
$N$ is proportional to the identity matrix and thus its matrix elements are
\begin{equation*}
  \langle F, 0 | N | F', 0 \rangle = N \delta_{F F'} .
\end{equation*}
From the Wigner-Eckart theorem one finds, that the matrix elements $\langle F', M' | T_0^{(2)} | F, M \rangle$ vanish for $|F - F'| > 2$. The nonzero matrix elements of $T_0^{(2)}$ are
\begin{equation*}
  \langle F, 0 | T_0^{(2)} | F, 0 \rangle \quad \text{and} \quad \langle F + 2, 0 | T_0^{(2)} | F, 0 \rangle .
\end{equation*}
The Wigner-Eckart theorem allows one to calculate the matrix elements of $T_0^{(2)}$ from a special class of matrix elements
\begin{eqnarray} \label{App-theta-Eq-4}
  & & \langle F + q, 0 | T_0^{(2)} | F, 0 \rangle = \frac{\langle F, 2; 0, 0 | F + q, 0 \rangle}{\langle F, 2; -F, -q | F + q, -F - q \rangle} \nonumber \\
  & & \mspace{50mu} \times \langle F + q, -F - q |T^{(2)}_{-q} | F, -F \rangle
\end{eqnarray}
for $q = 0, 2$. The brackets $\langle f_1, f_2; m_1, m_2 | f', m' \rangle$ are the Clebsch-Gordan coefficients (CGC). Eq.~(\ref{App-theta-Eq-4}) simplifies the calculation, since the states $|F, -F \rangle$ have a rather simple representation in the occupation number basis
\begin{equation} \label{App-theta-Eq-5}
  | F, -F \rangle = c_F \bigl( a_-^\dagger \bigr)^F \Bigl[ a_+^\dagger a_-^\dagger - \bigl( a_0^\dagger \bigr)^2 / 2 \Bigr]^{\frac{N - F}{2}} | 0, 0, 0 \rangle
\end{equation}
with the normalization constant
\begin{equation} \label{App-theta-Eq-6}
  \frac{1}{c_F^2} = \sum_{k = 0}^{\frac{N - F}{2}}\binom{\frac{N - F}{2}}{k}^2 \frac{k! (k + F)! (N - F - 2 k)!}{2^{N - F - 2 k}} .
\end{equation}
After inserting Eqs. (\ref{App-theta-Eq-3a}), (\ref{App-theta-Eq-3b}), (\ref{App-theta-Eq-5}) and the required CGCs into Eq.~(\ref{App-theta-Eq-4}), a lengthy calculation leads to the nonzero matrix elements
\begin{equation} \label{App-theta-Eq-7}
  \langle F, 0 | T_0^{(2)} | F, 0 \rangle = \frac{1}{\sqrt{6}} \biggl( 3 F - 2 N + 6 \frac{c_F^2}{d_F^2} \biggr) \frac{F + 1}{2 F - 1}
\end{equation}
and
\begin{eqnarray}
  & & \langle F + 2, 0 |T_0^{(2)} | F, 0 \rangle = \nonumber \\
  & & \mspace{30mu} \sqrt{\frac{3}{2}} \frac{c_F}{c_{F + 2}} \frac{N - F}{2} \sqrt{\frac{(F + 1)(F + 2)}{(2 F + 1)(2 F + 3)}} , \qquad
\end{eqnarray}
where
\begin{equation}
  \frac{1}{d_F^2} = \sum_{k = 0}^{\frac{N - F}{2}} k \binom{\frac{N - F}{2}}{k}^2 \frac{k!(k + F)!(N - F - 2 k)!}{2^{N - F - 2 k}} .
\end{equation}

The coefficients $\chi_F$ of the expansion (\ref{App-theta-Eq-1}) can be calculated by means of Eq.~(\ref{App-theta-Eq-5}). Using
\begin{equation}
  | F, 0 \rangle = 1 / \sqrt{(2 F)!} \bigl( F_+ \bigr)^F | F, -F \rangle
\end{equation}
we obtain
\begin{eqnarray} \label{App-theta-Eq-8}
  \chi_F & = & \langle F, 0 | 0, N, 0 \rangle = \langle F, -F | \bigl( F_- \bigr)^F | 0, N, 0 \rangle / \sqrt{(2 F)!} \nonumber \\
  & = & c_F \biggl( \! -\frac{1}{2} \biggr)^{\frac{N - 2 F}{2}} \sqrt{N! \binom{2 F}{F}^{-1}} .
\end{eqnarray}
The evolution is obtained by inserting Eqs.~(\ref{App-theta-Eq-7})--(\ref{App-theta-Eq-8}) into Eq.~(\ref{App-theta-Eq-2}). Only terms with $|F - F'| = 2$ lead to non-constant contributions to the evolution:
\begin{equation*}
  n_0(t)' = \sum_{F = F_\text{min}, \Delta F = 2}^{N-2} C_F \cos \bigl[ g_s (2 F + 3) t \bigr] ,
\end{equation*}
with frequencies $g_s (2 F + 3) = \omega_{F + 2} - \omega_F$ and amplitudes
\begin{eqnarray} \label{App-theta-Eq-9}
  & & C_F = (N - F) c_F^2 (N - 1)! \Bigl( \frac{1}{2} \Bigr)^{N - 2 F} \nonumber \\
  & & \times \sqrt{\binom{2 F}{F}^{-1} \binom{2 F + 4}{F + 2}^{-1} \frac{(F + 1)(F + 2)}{(2 F + 1)(2 F + 3)}} . \qquad
\end{eqnarray}

\begin{figure}[t]
  \includegraphics[width = \columnwidth]{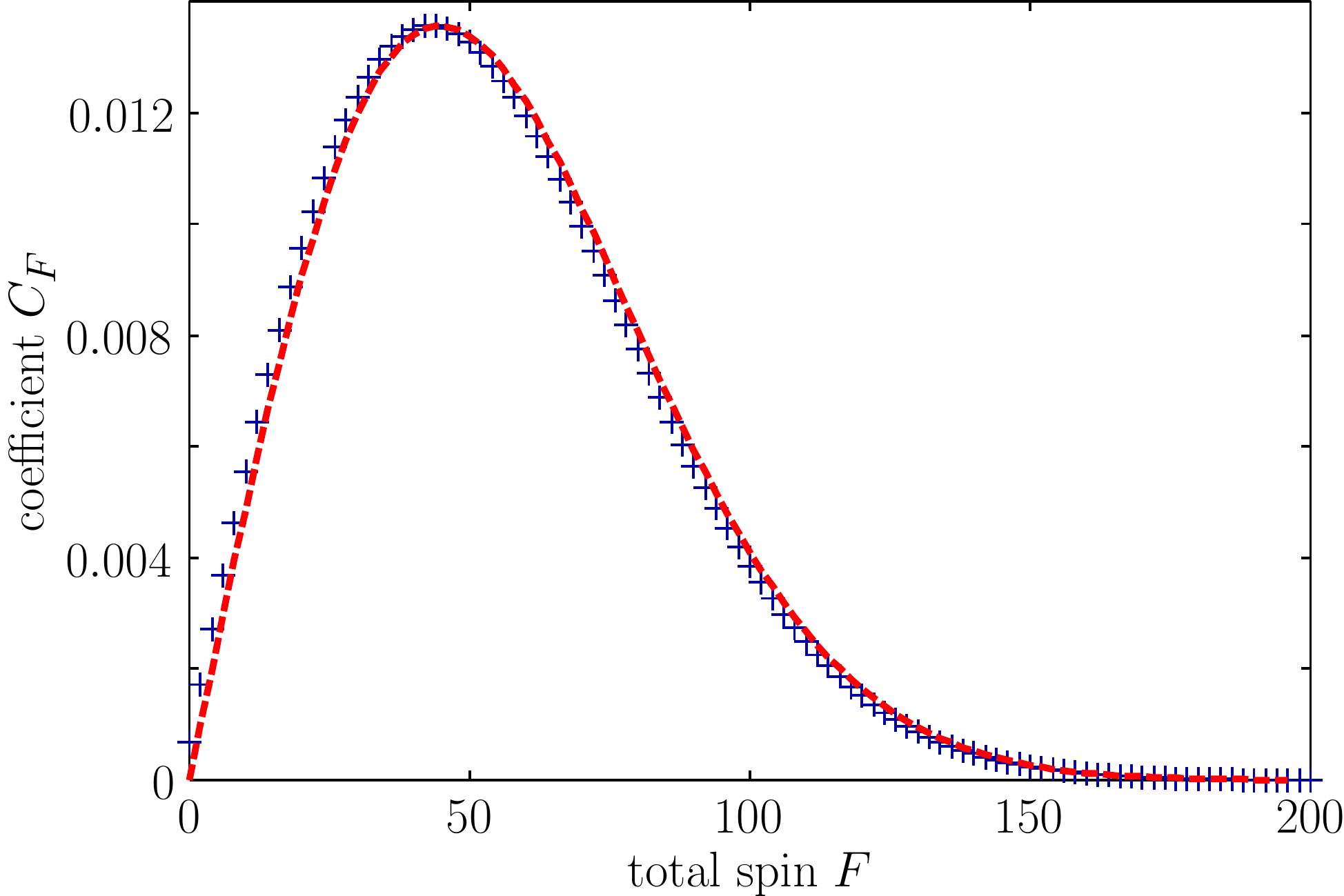}
  \caption{(color online). Exact amplitudes $C_F$ [(\ref{App-theta-Eq-9}), blue crosses], of 2000 atoms compared to the approximation [(\ref{App-theta-Eq-18}), red line].}
  \label{Fig-coefficient-distribution}
\end{figure}

The amplitudes (\ref{App-theta-Eq-9}) can be approximated for large $N$. The most involved part is the approximation of $c_F$. The factorials in (\ref{App-theta-Eq-6}) can be written as binomial coefficients.
\begin{equation*}
  k!(k + F)!(N - F - 2 k)! = N! \left( \binom{N}{\vartheta} \binom{\vartheta}{\frac{\vartheta - F}{2}} \right)^{-1} ,
\end{equation*}
where $\vartheta = 2 k + F$. We obtain from (\ref{App-theta-Eq-6})
\begin{equation} \label{App-theta-Eq-10}
  \frac{1}{c_F^2} = \sum_{\vartheta = F, \Delta \vartheta = 2}^N \frac{N!}{2^{N - \vartheta}} \binom{\frac{N - F}{2}}{\frac{\vartheta - F}{2}}^2 \left( \binom{N}{\vartheta} \binom{\vartheta}{\frac{\vartheta - F}{2}} \right)^{-1} .
\end{equation}
The terms with $\vartheta \approx N$ dominate the sum in (\ref{App-theta-Eq-10}). The approximation
\begin{equation} \label{App-theta-Eq-11}
  \frac{1}{2^n} \binom{n}{k} \approx \frac{1}{\sqrt{2 \pi} \sqrt{n / 4}} \exp \left[ -\frac{1}{2} \frac{(k - n / 2)^2}{n / 4} \right]
\end{equation}
holds for large $n$. Applying this to $\binom{\vartheta}{(\vartheta - F) / 2}$ gives
\begin{equation} \label{App-theta-Eq-12}
  \frac{1}{c_F^2} \approx \sum_{\vartheta = F}^N \frac{N!}{2^N} \sqrt{\frac{\pi \vartheta}{2}} \binom{\frac{N - F}{2}}{\frac{\vartheta - F}{2}}^2 \binom{N}{\vartheta}^{-1} \exp\left( \frac{F^2}{2 \vartheta} \right) .
\end{equation}
Further, we approximate $\vartheta = N$ in the exponential and the square root of (\ref{App-theta-Eq-12}). The exponential dependence on $F^2$ shows, that $c_F$ is negligible for $F \gg N$. Thus, we assume $F \ll N$ in the following. The binomials in (\ref{App-theta-Eq-12}) are expanded into factorials and approximated using $(n - k)!\approx n! / n^k$ for $n \gg k$:
\begin{equation} \label{App-theta-Eq-13}
  \binom{\frac{N - F}{2}}{\frac{\vartheta - F}{2}}^2 \binom{N}{\vartheta}^{-1} \approx \left( \frac{N - F}{2 N} \right)^{N-\vartheta}  \binom{N - \vartheta}{\frac{N - \vartheta}{2}} .
\end{equation}
We apply (\ref{App-theta-Eq-11}) to the right-hand side of (\ref{App-theta-Eq-13})
\begin{equation}
  \Bigl( \frac{1}{2} \Bigr)^{N - \vartheta} \binom{N - \vartheta}{\frac{N - \vartheta}{2}} \approx \sqrt{\frac{2}{\pi (N - \vartheta)}} .
\end{equation}
Further, for $F \ll N$, one gets to first order
\begin{equation} \label{App-theta-Eq-14}
  \Bigl( \frac{N - F}{N} \Bigr)^{N - \vartheta} \approx \exp \left( -\frac{F (N - \vartheta)}{N} \right) .
\end{equation}
We insert the approximations (\ref{App-theta-Eq-13})--(\ref{App-theta-Eq-14}) into (\ref{App-theta-Eq-12}) and approximate the sum by an integral
\begin{equation*}
  \frac{1}{c_F^2} \approx  e^{-F}e^{F^2 / (2N)} \frac{N!}{2^N} \frac{1}{2} \int_F^N d \vartheta \sqrt{\frac{N}{N - \vartheta}} \medspace e^{F \frac{\vartheta}{N}} ,
\end{equation*}
where an additional factor $1 / 2$ accounts for the step size of 2 in the sum. The integral has the value
\begin{equation} \label{App-theta-Eq-15}
\int_F^N d \vartheta \sqrt{\frac{N}{N - \vartheta}} \medspace e^{F \frac{\vartheta}{N}} = -N \sqrt{\frac{\pi}{F}} \medspace e^{F} \text{erf} \left( \sqrt{\frac{F (N - \vartheta)}{N}} \right)
\end{equation}
with the error function erf$(x)$. Inserting the upper limit $N$ of the integral into (\ref{App-theta-Eq-15}) leads to $\text{erf} (0) = 0$. With $F \ll N$, the lower limit $F$ gives $\text{erf} (\sqrt{F}) \approx 1$ for $F \geq 5$ and thus
\begin{equation} \label{App-theta-Eq-16}
  \frac{1}{c_F^2} \approx \frac{N!}{2^N} \frac{\sqrt{\pi}}{2} \frac{N}{\sqrt{F}} \exp \left( \frac{1}{2} \frac{F^2}{N} \right) .
\end{equation}
The maximum of (\ref{App-theta-Eq-16}) is at $F = \sqrt{N/2}$. The Gaussian $\exp (-F^2 / N)$ has a width of $\sqrt{N / 2}$, which is an upper bound for the width of $c_F$. Now we approximate the other terms in (\ref{App-theta-Eq-9}). Since $F \approx \sqrt{N}$ for the dominant contributions,
\begin{equation}
  \sqrt{\binom{2 F}{F} \binom{2 F + 4}{F + 2}} \approx \binom{2 F}{F}\approx 2^{2 F} \sqrt{\frac{1}{\pi F}} ,
\end{equation}
where we used (\ref{App-theta-Eq-11}) in the last step. For $1 \ll F \ll N$, the remaining factors become
\begin{equation} \label{App-theta-Eq-17}
  \frac{N - F}{N} \cdot \sqrt{\frac{(F + 1) (F + 2)}{(2 F + 1) (2 F + 3)}} \approx \frac{1}{2} .
\end{equation}
Inserting (\ref{App-theta-Eq-16})--(\ref{App-theta-Eq-17}) into (\ref{App-theta-Eq-9}) leads to
\begin{equation} \label{App-theta-Eq-18}
 C_F \approx \frac{F}{N} \exp \left(-\frac{1}{2} \frac{F^2}{N} \right) .
\end{equation}
The approximation (\ref{App-theta-Eq-18}) and the exact amplitude (\ref{App-theta-Eq-9}) are compared for $2000$ atoms in Fig.~\ref{Fig-coefficient-distribution}.

\end{appendix}


\begin{thebibliography}{22}

\bibitem{Stenger98} J. Stenger, S. Inouye, D. M. Stamper-Kurn, H.-J. Miesner, A. P. Chikkatur, and W. Ketterle, \doi{10.1038/24567}{Nature {\bf 396}, 345 (1998).}

\bibitem{Sadler06} L. E. Sadler, J. M. Higbie, S. R. Leslie, M. Vengalattore, and D. M. Stamper-Kurn, \doi{10.1038/nature05094}{Nature {\bf 443}, 312 (2006).}

\bibitem{Vengalattore08} M. Vengalattore, S. R. Leslie, J. Guzman, and D. M. Stamper-Kurn, \doi{10.1103/PhysRevLett.100.170403}{Phys. Rev. Lett. {\bf 100}, 170403 (2008).}

\bibitem{Vengalattore09} M. Vengalattore, J. Guzman, S. R. Leslie, F. Serwane, and D. M. Stamper-Kurn, \arxiv{0901.3800}{arXiv:0901.3800.}

\bibitem{Kronjaeger09} J. Kronj\"ager, C. Becker, P. Soltan-Panahi, K. Bongs, and K. Sengstock, \arxiv{0904.2339}{arXiv:0904.2339.}

\bibitem{Lamacraft07} A. Lamacraft, \doi{10.1103/PhysRevLett.98.160404}{Phys. Rev. Lett. {\bf 98}, 160404 (2007).}

\bibitem{Ho98} T.-L. Ho, \doi{10.1103/PhysRevLett.81.742}{Phys. Rev. Lett. {\bf 81}, 742 (1998).}

\bibitem{Ohmi98} T. Ohmi and K. Machida, \doi{10.1143/JPSJ.67.1822}{J. Phys. Soc. Jpn. {\bf 67}, 1822 (1998).}

\bibitem{Law98} C. K. Law, H. Pu, and N. P. Bigelow, \doi{10.1103/PhysRevLett.81.5257}{Phys. Rev. Lett. {\bf 81}, 5257 (1998).}

\bibitem{Koashi00} M. Koashi and M. Ueda, \doi{10.1103/PhysRevLett.84.1066}{Phys. Rev. Lett. {\bf 84}, 1066 (2000).}

\bibitem{Schmaljohann04} H. Schmaljohann, M. Erhard, J. Kronj\"ager, M. Kottke, S. van Staa, L. Cacciapuoti, J. J. Arlt, K. Bongs, and K. Sengstock, \doi{10.1103/PhysRevLett.92.040402}{Phys. Rev. Lett. {\bf 92}, 040402 (2004).}

\bibitem{Chang04} M.-S. Chang, C. D. Hamley, M. D. Barrett, J. A. Sauer, K. M. Fortier, W. Zhang, L. You, and M. S. Chapman, \doi{10.1103/PhysRevLett.92.140403}{Phys. Rev. Lett. {\bf 92}, 140403 (2004).}

\bibitem{Kuwamoto04} T. Kuwamoto, K. Araki,T.  Eno, and T. Hirano, \doi{10.1103/PhysRevA.69.063604}{Phys. Rev. A {\bf 69}, 063604 (2004).}

\bibitem{Widera05} A. Widera, F. Gerbier, S. F\"olling, T. Gericke, O. Mandel, and I. Bloch, \doi{10.1103/PhysRevLett.95.190405}{Phys. Rev. Lett. {\bf 95}, 190405 (2005).}

\bibitem{Widera06} A. Widera, F. Gerbier, S. F\"olling, T. Gericke, O. Mandel, and I. Bloch, \doi{10.1088/1367-2630/8/8/152}{New. J. Phys. {\bf 8}, 152 (2006).}

\bibitem{Chang05} M.-S. Chang, Q. Qin, W. Zhang, L. You, and M. S. Chapman, \doi{10.1038/nphys153}{Nature Phys. {\bf 1}, 111 (2005).}

\bibitem{Zhang05} W. Zhang, D. L. Zhou, M.-S. Chang, M. S. Chapman, and L. You, \doi{10.1103/PhysRevA.72.013602}{Phys. Rev. A {\bf 72}, 013602 (2005).}

\bibitem{Kronjaeger06} J. Kronj\"ager, C. Becker, P. Navez, K. Bongs, and K. Sengstock, \doi{10.1103/PhysRevLett.97.110404}{Phys. Rev. Lett. {\bf 97}, 110404 (2006).}

\bibitem{Black07} A. T. Black, E. Gomez, L. D. Turner, S. Jung, and P. D. Lett, \doi{10.1103/PhysRevLett.99.070403}{Phys. Rev. Lett. {\bf 99}, 070403 (2007).}

\bibitem{Goldstein99} E. V. Goldstein and P. Meystre, \doi{10.1103/PhysRevA.59.3896}{Phys. Rev. A {\bf 59}, 3896 (1999).}

\bibitem{Leslie09} S. R. Leslie, J. Guzman, M. Vengalattore, J. D. Sau, M. L. Cohen, and D. M. Stamper-Kurn, \doi{10.1103/PhysRevA.79.043631}{Phys. Rev. A {\bf 79}, 043631 (2009).}

\bibitem{Klempt09} C. Klempt, O. Topic, G. Gebreyesus, M. Scherer, T. Henninger, P. Hyllus, W. Ertmer, L. Santos, and J. J. Arlt, \doi{10.1103/PhysRevLett.103.195302}{Phys. Rev. Lett. {\bf 103}, 195302 (2009).}

\bibitem{Klempt10} C. Klempt, O. Topic, G. Gebreyesus, M. Scherer, T. Henninger, P. Hyllus, W. Ertmer, L. Santos, and J. J. Arlt, \doi{10.1103/PhysRevLett.104.195303}{Phys. Rev. Lett. {\bf 104}, 195303 (2010).}

\bibitem{Sorensen01} A. S{\o}rensen, L.-M. Duan, J. I. Cirac, and P. Zoller, \doi{10.1038/35051038}{Nature {\bf 409}, 63 (2001).}

\bibitem{Pu00} H. Pu and P. Meystre, \doi{10.1103/PhysRevLett.85.3987}{Phys. Rev. Lett. {\bf 85}, 3987 (2000).}

\bibitem{Duan00} L.-M. Duan, A. S{\o}rensen, J. I. Cirac, and P. Zoller, \doi{10.1103/PhysRevLett.85.3991}{Phys. Rev. Lett. {\bf 85}, 3991 (2000).}

\bibitem{Duan02} L.-M. Duan, J. I. Cirac, and P. Zoller, \doi{10.1103/PhysRevA.65.033619}{Phys. Rev. A {\bf 65}, 033619, (2002).}

\bibitem{Muestecaplioglu02} \"O. E. M\"ustecapl{\i}o\u{g}lu, M. Zhang, and L. You, \doi{10.1103/PhysRevA.66.033611}{Phys. Rev. A {\bf 66}, 033611 (2002).}

\bibitem{Cui08} X. Cui, Y. Wang, and F. Zhou, \doi{10.1103/PhysRevA.78.050701}{Phys. Rev. A {\bf 78}, 050701(R) (2008).}

\bibitem{Lieb00} E. H. Lieb, R. Seiringer, and J. Yngvason, \doi{10.1103/PhysRevA.61.043602}{Phys. Rev. A {\bf 61}, 043602 (2000).}

\bibitem{Erdos07} L. Erd{\H{o}}s, B. Schlein, and H.-T. Yau, \doi{10.1103/PhysRevLett.98.040404}{Phys. Rev. Lett. {\bf 98}, 040404 (2007).}

\bibitem{Chen09} Z. Chen, C. Bao, and Z. Li, \doi{10.1143/JPSJ.78.114002}{J. Phys. Soc. Jpn. {\bf 78}, 114002 (2009).}

\bibitem{Eisenberg02} E. Eisenberg and E. H. Lieb, \doi{10.1103/PhysRevLett.89.220403}{Phys. Rev. Lett. {\bf 89}, 220403 (2002).}

\bibitem{Kronjaeger05} J. Kronj\"ager, C. Becker, M. Brinkmann, R. Walser, P. Navez, K. Bongs, and K. Sengstock, \doi{10.1103/PhysRevA.72.063619}{Phys. Rev. A {\bf 72}, 063619 (2005).}

\bibitem{Kronjaeger07} J. Kronj\"ager, Ph.D. thesis, University of Hamburg (2007); available online: \urn{urn:nbn:de:gbv:18-34281}{urn:nbn:de:gbv:18-34281}.

\bibitem{Yaffe82} L. G. Yaffe, \doi{10.1103/RevModPhys.54.407}{Rev. Mod. Phys. {\bf 54}, 407 (1982).}

\bibitem{Heinze09} J. Heinze, diploma thesis, University of Hamburg (2009); \web{http://www.physnet.uni-hamburg.de/hp/theorie/Diplomarbeiten/Jannes_Heinze_Diplomarbeit_11_02_2009.pdf}{available online.}

\bibitem{Kronjaeger08} J. Kronj\"ager, K. Sengstock, and K. Bongs, \doi{10.1088/1367-2630/10/4/045028}{New. J. Phys. {\bf 10}, 045028 (2008).}

\bibitem{Weiss08} C. Weiss and N. Teichmann, \doi{10.1103/PhysRevLett.100.140408}{Phys. Rev. Lett. {\bf 100}, 140408 (2008).}

\bibitem{Bronstein} I. N. Bronstein, K. A. Semendjajew, G. Musiol, and H. M\"uhlig, {\it Taschenbuch der Mathematik} (Verlag Harri Deutsch, 2001).

\end{thebibliography}
\end{document}